\documentclass[preprint,12pt,authoryear]{elsarticle}

\usepackage[english]{babel}
\usepackage[T1]{fontenc}
\usepackage[utf8]{inputenc}
\usepackage[normalem]{ulem}

\usepackage{amsfonts,amssymb,amsmath,amsthm, graphics, setspace}
\usepackage{bbm}
\usepackage{xcolor}
\usepackage{graphics}
\usepackage{mathtools}
\usepackage{hyperref} 
\usepackage{soul}
\usepackage[normalem]{ulem}
\usepackage{multirow}

\usepackage{lineno}

\journal{Theoretical Population Biology}

\begin{document}

\begin{frontmatter}

\author[nancy]{Juan Mardomingo-Sanz\corref{cor1}}
\ead{juan.mardomingo-sanz@univ-lorraine.fr}
\ead[url]{https://iecl.univ-lorraine.fr/membre-iecl/mardomingo-sanz-juan/}

\author[telecom]{Raphaël Tran-Thanh}

\author[inrae]{Camille Coron}

\ead[url]{https://camillecoron.github.io/}
\cortext[cor1]{Corresponding author}

\author[gqe]{Diala Abu Awad}
\ead[url]{https://moulon.inrae.fr/personnes/dabuawad/}

\affiliation[nancy]{organization={Université de Lorraine, CNRS, Inria, IECL},
            postcode={F-54000},
            city={Nancy},
            country={France}}
            
\affiliation[telecom]{organization={Institut Polytechnique de Paris, Telecom Paris, LTCI},
            postcode={91120}, 
            city={Palaiseau},
            country={France}}
            
\affiliation[inrae]{organization={Université Paris-Saclay, AgroParisTech, INRAE, UMR MIA Paris-Saclay},
            postcode={91120},
            city={Palaiseau},
            country={France}}
            
\affiliation[gqe]{organization={Université Paris-Saclay, INRAE, CNRS, AgroParisTech, GQE - Le Moulon},
            postcode={91190},
            city={Gif-sur-Yvette},
            country={France}}
            
\title{How timing and mode of selection shape ancestral genetic contributions}

\begin{abstract}
We examine a population that evolves according to a biparental Moran model with two distinct types of individuals. Each individual's type is determined by the allele they carry at one biallelic locus, which is transmitted in a Mendelian way. This type influences the individual's mating preferences. We investigate two specific mating patterns: (i) where the type determines the level of attractiveness, and (ii) where the type dictates the level of homogamous choosiness. In both scenarios, one type has an advantage over the other. Assuming an initial proportion $b$ of individuals are of the advantaged type, we give an explicit formula for the asymptotic proportion of the population's genome that originates from these advantaged ancestors under an infinite population size approximation. Our results enable a comparison of the relative strengths of homogamous choosiness, attractiveness, and viability selection in terms of genome transmission, and prove notably that viability selection is always more efficient than both forms of reproductive selection.
\end{abstract}


\end{frontmatter}

\newtheorem{theorem}{Theorem}[section]
\newtheorem{corollary}[theorem]{Corollary}
\newtheorem{proposition}[theorem]{Proposition}

\newcommand{\camille}[1]{\textcolor{blue}{#1}}
\newcommand{\raphael}[1]{\textcolor{red}{#1}}
\newcommand{\juan}[1]{\textcolor{cyan}{#1}}
\newcommand{\diala}[1]{\textcolor{magenta}{#1}}

\newcommand{\commentaire}[1]{\marginpar{\raggedright\footnotesize\color{blue}#1}}

\section{Introduction}\label{sec:context}

The role of selection in shaping the genetic structure and diversity of populations has been central to population genetics (\cite{Kimura1962,KimuraOhta1969}). Initial approaches examined selection at a single locus, but it became evident that the consequences of selection spread beyond the focal site and could influence genetic diversity throughout the genome. As a consequence of physical linkage, alleles under positive (resp. negative) selection tend to decrease genetic diversity around the focal locus as they go to fixation (resp. are purged), resulting in a selective sweep \citep{smithhaigh1974, Stephanetal1992} (resp. background selection, \cite{Charlesworthetal1995}). While most works have addressed local changes in genetic diversity during sweeps, it is plausible that the decrease in genetic diversity may extend beyond that genomic region. Using forward-in-time simulations, \cite{Ollivieretal2025} showed that populations with asexual and sexual phases of reproduction show a genome-wide decrease in genetic diversity following a selective sweep, \textit{i.e.} including chromosomes that are not under selection.

Understanding how selection modes shape genomic diversity has direct implications for empirical inference. When interpreting genomic scans for selection, distinguishing between viability-based and mating-preference-driven sweeps could improve our ability to identify the nature of past selective events. Just as distinguishing hard from soft sweeps \citep{HermissonPennings2017} transformed the interpretation of genomic scans \citep{SchriderKern2016}, distinguishing selection modes could similarly refine inference. Hermaphroditic species—such as many plants, invertebrates, and fungi—are particularly relevant, as they experience diverse selection pressures, notably on mating \citep{Beekmanetal2016}, that may leave distinct genealogical signatures. Yet it remains unclear whether different widespread modes of selection produce measurably different genome-wide consequences.

Theoretical approaches to quantify genetic diversity dynamics during sweeps have mostly focused on local changes by examining single sweeps, or more genome-wide effects by introducing consecutive sweeps \citep{Stephanetal1992, Leocard_2009, HartfieldBataillon2020, Marshetal2026}, using a variety of approaches. But beyond the local effects of a single sweep, how does a strong selective event affect the transmission of genetic material initially present in the parental genome carrying the mutant beneficial allele?

The biparental Moran model proposed by \cite{CoronLeJan2022JMB} provides a natural framework to study this question rigorously, without having to resort to following dynamics at many loci. Unlike Wright-Fisher approximations that can obscure pedigree-level transmission, this framework generates an explicit pedigree on which gene genealogies are backwards-in-time random walks. The genetic contribution of an ancestor is the probability that a gene sampled uniformly in the current population at a locus independent from the one under selection descends from that individual. \cite{coronLeJan2024,coronLeJan2025} have characterised these dynamics under viability selection, where an allele at the focal locus reduces the death rate. In a related theoretical framework the impact of self-fertilisation on genome transmission has also been studied by \cite{Newman2024}.

The strength of a sweep on the genome level can be defined as the efficiency of transmission of ancestral genomic material, that is the genetic diversity initially associated with the favourable allele. A natural measure of this efficiency is the genetic weight of the initially advantaged ancestors, \textit{i.e.} the asymptotic fraction of the population's genome that descends from those ancestors as the advantaged type fixes (\cite{BartonEtheridge2011,Newman2024}). This quantity goes beyond the allele frequency trajectory: two forms of selection may drive allele frequencies at identical speed yet leave markedly different genealogical footprints, and the genetic weight captures precisely this difference. Such footprints are closely related to the notion of genetic hitchhiking and to the width of selective sweeps in molecular evolution, and are therefore of direct interest for developing statistical tools to detect and classify selection from genomic data.

\cite{CoronLeJan2022JMB} show that once an allele increasing viability has gone to fixation, up to $19\%$ of the genome of individuals present in the population will be that initially associated with that allele if $1\%$ of the initial population carried it. Though selection for viability is an important measure of fitness, natural selection comprises many different forms of selection. In sexually reproducing populations, fitness depends on several components, including mating preferences. Selection on whom an individual chooses to mate with will violate the assumption of panmictic mating (\cite{Carson2003}) and hence will impact how genetic material is transmitted from one generation to the next. It is well known that non-random mating shapes the spread of advantageous alleles and leaves a distinctive genealogical signature in the population \citep{Newman2024,BartonEtheridge2011,smithhaigh1974}. As many species of interest do reproduce sexually, it is pertinent to question whether selection on mating preference leads to different consequences of selective sweeps on the genome-wide level compared to viability selection. This has, to our knowledge, not yet been examined.

Mate choice can exist in different forms and at different scales, from selection on behaviour down to selection on gametes. Attractiveness (often of males, \cite{Pilastroetal2004, Waeltietal2009, Draytonetal2010, Wainwrightetal2016}) and choosiness (more commonly applied by females, \cite{Bleuetal2012, YukilevichAoki2022, XuZuk2025, Yukilevich2025}) are known to evolve as a response to selection, increasing access to reproduction in the former and, hopefully, to better offspring in the latter. Though mate selection is mainly applied to species with separate sexes, there is ample proof that hermaphrodites can also undergo selection that will affect investment and expression of their female and male function \citep{Leonard2006, Beekmanetal2016}. These pressures of selection likely contribute to shaping the genome-wide patterns of diversity following selective sweeps, but may do so differently than viability selection, due to the inherent asymmetry in the selection for mating preferences.

In this work, we focus on genome-wide effects, , under the assumption that there is no physical linkage along the genome, focusing on the contribution of selection mode to transmission efficiency. To address how different widespread modes of selection influence the genome-wide consequences of a selective sweep at a locus under positive selection, we compare selection for attractiveness, homogamous choosiness and on viability. Our main findings are that viability selection always leads to a strictly higher asymptotic genetic contribution, meaning that genome-wide diversity is reduced to a greater degree than if selection was on mating preference. We also show that selection through increased attractiveness leads to a faster invasion of the advantaged type and a larger asymptotic genetic contribution of the initially advantaged ancestors than selection on homogamous choosiness, and this is true for any given selection parameter and any initial proportion of the advantaged type. Intuitively, this is because increased attractiveness biases every mating event in the population in favour of the advantaged type, whereas homogamous choosiness only operates when choosy individuals are chosen to mate. The form of the mating preference matters most in the regime of moderate selection (arguably the most relevant one for natural populations).

\section{Model}\label{sec:model}

\paragraph{General model} 
We model the genetic dynamics of a population using a bi-type biparental Moran model with selection, partly following the framework of previous papers \cite{CoronLeJan2022JMB,CoronLeJan2022UMJ,coronLeJan2024,coronLeJan2025}. The present article, however, focuses on different forms of reproductive selection rather than viability selection. More precisely, we consider hermaphroditic and haploid individuals characterised by their genome at one biallelic locus. Thus, at this locus, individuals carry one allele inherited from one of their parents chosen uniformly, and are said to be either of type $0$ or of type $1$, depending on this allele.

\begin{figure}
    \centering
    \includegraphics[scale=0.14]{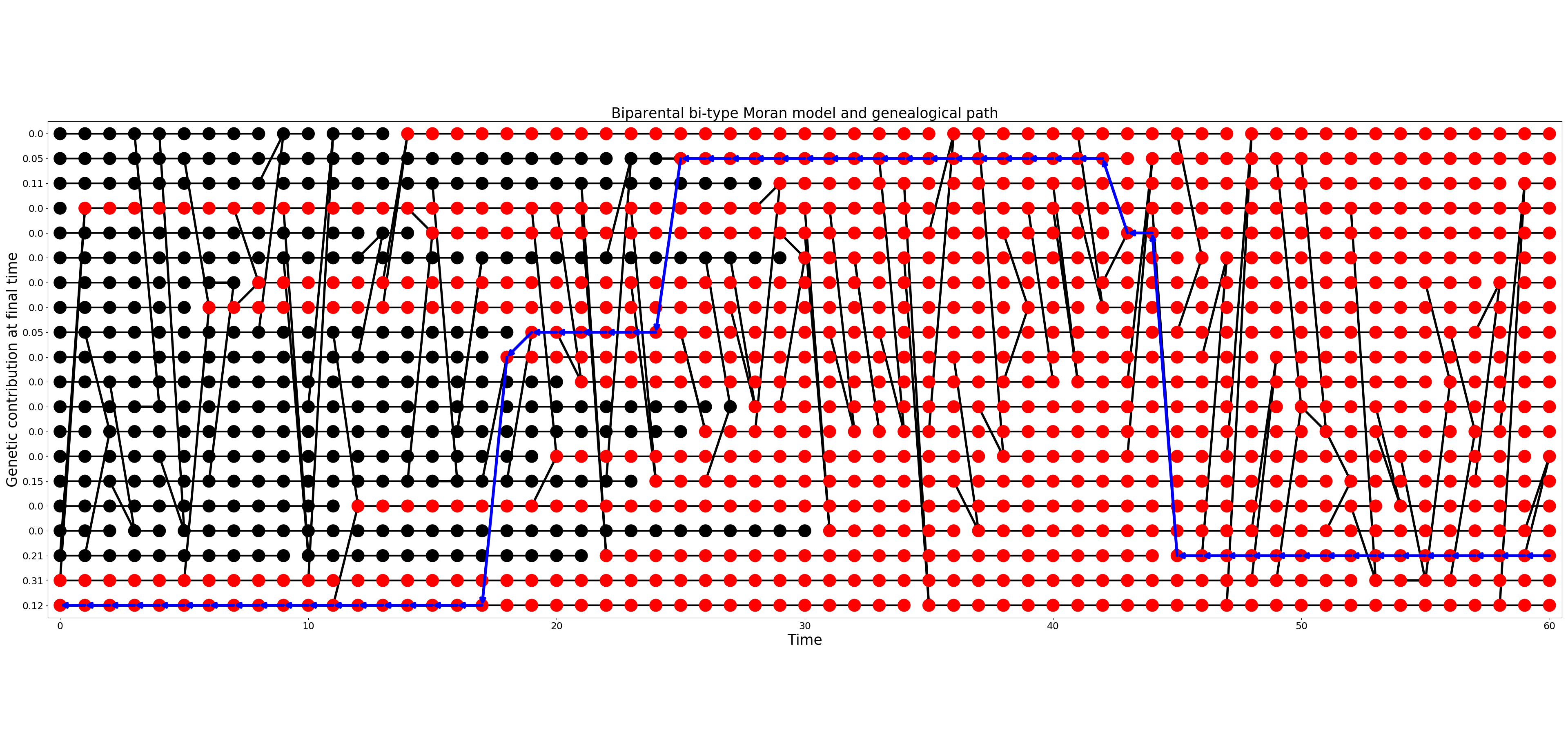}
    \caption{
Pedigree in the two-type biparental Moran model with $N=20$ individuals over $60$ time steps and very strong selection. Types are represented by colors (black or red). The blue path represents the genealogical history of a gene sampled uniformly in the population at time $n=60$. Numbers in ordinate give probabilities that a gene sampled uniformly at time $60$ comes from each of the initial individuals.}
\label{fig-Pedigree}
\end{figure}

At each discrete time step $n\in \mathbb{Z}_+$, individuals occupy sites indexed by $I=\{1,2,...,N\}$. We start by picking a first parent, called mother and denoted by $\mu_n \in I$, uniformly at random among the population. Then, according to the mating preferences induced by its type, the mother chooses a partner, called father, denoted by $\pi_n \in I$. More specifically, we introduce the mating preference parameters $s_0$ and $s_1$ and assume that if the mother is of type $i \in \{0, 1\}$, a given individual of type $1$ is $1+s_i$ times more likely to be chosen as mating partner than a given individual of type $0$. Finally, one of the two parents, denoted $\nu_n \in \{ \mu_n, \pi_n \}$, is chosen with equal probability to pass on its type to their unique offspring. The offspring finally replaces an individual $\kappa_n \in I$ chosen uniformly among the population, so that the population size remains constant. This forward-in-time dynamics naturally yields the pedigree $G$ of the population, defined as the set of parental links between individuals, or the support of genetic material transmission, traced backwards in time. We denote by $G_n$ the pedigree until time $n$. An illustration of the pedigree, and types of individuals is given in Figure \ref{fig-Pedigree}.

\medskip

\paragraph{Two particular cases of mating preferences}
Throughout the article, we will focus on two particular mating patterns inducing reproductive selection. In the first case, called \textit{increased attractiveness}, we assume that all mothers, regardless of their own type, preferentially choose a partner of type $1$, so that type $1$ individuals are, as fathers, uniformly more attractive than type $0$. Formally, this corresponds to setting $s_0=s_1=s>0$. Note that this modeling can also refer to a difference in fertility of fathers (or mothers) and a uniform choice of their partner. However this interpretation is less natural since fertility is generally expected to depend on both parents' types. In the second case, called \textit{homogamous choosiness}, we assume that mothers with type $0$ choose their partner uniformly without showing any preference, whereas mothers with type $1$ exhibit homogamous behaviour, i.e. preferably mate with type $1$ individuals. This scenario corresponds to setting $s_0=0$ and $s_1>0$. In both cases, individuals with type $1$ are advantaged compared to those with type $0$.

Selection in monoparental (or clonal) populations modeled by the classical Moran model has been studied in several articles, notably \cite{Cordero2017, KluthBaake} that focus on reproductive selection, but lead to different limiting deterministic approximations, mainly because in our model the first parent is chosen uniformly in the population, i.e. regardless of its type.

\paragraph{Ancestor genetic weights}
We are interested in studying the impact of mating preferences on individual genome transmission. To this aim, as in \cite{coronLeJan2025}, we consider a second locus, distant enough from the one which is under selection so that the genome can be assumed to transmit independently at these two loci. We then sample a gene at this second locus in an individual among the population at time $n$, and consider the genealogy of this gene, i.e. the sequence of ancestral individuals by which this gene was transmitted, at all times $n-k\leq n$. This genealogy, denoted by $(X^{(n)}_k, n-k)_{0\leq k\leq n}$, is a random walk, backward in time on the pedigree $G_n$. An example of realization of such a random walk is given by the blue path in Figure \ref{fig-Pedigree}. Now an interesting quantity is the probability $W_n(i,j)=\mathbb{P}(X^{(n)}_0=j|X^{(n)}_n=i, G_n)$ that a gene sampled from the individual $i$ in the population at time $n$ originates from the ancestor $j$ at time $0$, as a function of the pedigree until time $n$. This quantity can be defined as the genetic weight of ancestor $j$ in the individual $i$. Indeed, if the genome is composed of infinitely many independent loci with equal size, then this quantity will be equal to the proportion of the genome of individual $i$ living at time $n$, coming from ancestor $j$ living at time $0$, given the pedigree. As an example, we give in the y-axis of Figure \ref{fig-Pedigree} the means $\frac{1}{N}\sum_{i}W_n(i,j)$ for all $j\in I$, that give the probabilities that a gene sampled uniformly at time $n$ comes from each of the ancestors living at time $0$, knowing the pedigree.

Let us denote by $\mathcal{Y}_n$ the set of individuals of type $1$ (advantaged) at time $n$ and by $Y_n= |\mathcal{Y}_n|$ its cardinality. As in \cite{coronLeJan2025}, the following two quantities will play a significant role in the study of genetic transmission within the population:
\[U_n := \sum_{l \in \mathcal{Y}_n} \sum_{l' \in \mathcal{Y}_0} W_n(l, l') \quad\quad\text{and}\quad\quad V_n := \sum_{l \notin \mathcal{Y}_n} \sum_{l' \in \mathcal{Y}_0} W_n(l, l').\]
The quantity $U_n$ (resp. $V_n$), taking values in $[0, N]$, denotes the amount of genome carried by advantaged (resp. disadvantaged) individuals living at time $n$, that comes from the initially advantaged ancestors. In particular, $U_0=Y_0$ and $V_0=0$. As long as $0<Y_n<N$, the normalised quantities
\[
\dfrac{U_n}{Y_n} \quad \text{and} \quad \frac{V_n}{N-Y_n} \in [0, 1]
\]
respectively give the probability, given the pedigree until time $n$, $G_n$, that a gene sampled from an advantaged (resp. disadvantaged) individual at time $n$ originates from an initially advantaged individual.

\section{Results and biological interpretation}\label{sec:results}

For any of the considered mating patterns, let us introduce the rescaled process $$(Z_n)_{n\in \mathbb{Z}_+} = \Big( \frac{Y_n}{N},~\frac{U_n}{N},~\frac{V_n}{N}\Big)_{n\in \mathbb{Z}_+}.$$ For any $n\geq0$, $Z_n$ takes values in $[0,1]^3$, and $Z_0=(Y_0/N,Y_0/N,0)$. Note that $Z_n$ also depends on the population size $N$, although we will not specify this dependence for the sake of simplicity. We denote by $\|.\|$ the Euclidean norm on $\mathbb{R}^3$ and by $(\mathcal{F}_n^Z,n\in\mathbb{Z}_+)$ the natural filtration associated to the stochastic process $(Z_n)_{n\in \mathbb{Z}_+}$. The process $(Z_n)_{n\in \mathbb{Z}_+}$ is not a Markov chain, because the dynamics of the genetic weights $(U_n)_{n\in\mathbb{Z}_+}$ and $(V_n)_{n\in\mathbb{Z}_+}$ between time $n$ and time $n+1$ depends on the exact genetic weights carried by all individuals, determined by the whole trajectory of $(\pi_k, \mu_k, \kappa_k)_{k\leq n}$, and not only by the current mean weights and number of advantaged individuals given by $Z_n$. The dynamics of the process $(Z_n)_{n\in \mathbb{Z}_+}$ is described in Section \ref{sec-Dyn-Z-app}. Our work focuses on two distinct forms of mating preferences: we first treat them separately in Subsections \ref{sec-attract} and \ref{sec-homo} and next compare them in Section \ref{sec-comparison}. These two forms of selection are also compared to viability selection, introduced in Section \ref{sec-viability}. As each generation of the Moran model changes the allele frequency of the population only by $1/N$, we study here the asymptotic behaviour of the stochastic process when the population size $N$ tends to infinity at a timescale of the order of $N$ to obtain a deterministic approximation, as it has been done in \cite{coronLeJan2025}.

\subsection{Increased attractiveness}\label{sec-attract}
In this section we focus on the increased attractiveness case. The following theorem, whose proof is postponed to \ref{sec-attract-app}, gives the asymptotic behaviour of the stochastic process $(Z_n)_{n\in\mathbb{Z}_+}$.
\begin{theorem} \label{Convergence_pref} Let $b \in (0,1)$. If the initial proportion of advantaged individuals $\frac{Y_0}{N}$ converges to $b\in (0,1)$  in probability when $N$ goes to infinity, then for any $c \in \mathbb{R}_{+}$, 
$$ \sup_{\, 0\leq t \leq c} \| Z_{\lfloor Nt \rfloor} - z_a(t) \| \xrightarrow[N \rightarrow \infty]{} 0,$$ in probability, where $(z_a(t),t\geq0)=((y_a(t),u_a(t),v_a(t)),t\geq0)$ is the unique solution of the following differential system: 
\begin{equation}\label{eqdiff_pref}
    \begin{cases}
      y' = \dfrac{(s/2)y(1-y)}{1+sy} \\[10pt]
     u' = -\dfrac{u(1-y)}{1+sy}+ \left(\dfrac{2+s}{4}\right)\dfrac{vy + (1-y)u}{1+sy}\\[10pt]
     v' = -\dfrac{vy(1+s)}{1+sy} + \left(\dfrac{2+s}{4}\right)\dfrac{vy+ (1-y)u}{1+sy},
    \end{cases}
\end{equation}
with initial condition $(b,b,0)$.
\end{theorem}

The subscript "$a$" stands for "attractiveness".
The proofs of this theorem and of the analogous Theorem \ref{Convergence_homo} presented in Section \ref{sec-homo} show that it is natural to transform the vector $Z=(Y/N,U/N,V/N)\in[0,1]^3$ by a change of variable given by $(Y/N, \Delta, (U+V)/N)\in[0,1]\times[-1,1]\times[0,1]$, where $\Delta=\frac{U}{Y}-\frac{V}{1-Y}$ for any $Y\in(0,1)$. The quantity $\Delta$ is the difference between the average quantities of genome coming from advantaged ancestors that are carried by advantaged and disadvantaged individuals respectively. This difference can be seen as a shift from an equilibrium in which the genome of ancestors is well distributed among all individuals, analogous to linkage disequilibrium, or the shift from Hardy-Weinberg equilibrium. We expect this difference to go to $0$, because the quantity of transmitted genome by parents to offspring at each reproduction does not depend on the type of the newborn individual. The sum $(U+V)/N$ is simply the total proportion of genetic material coming from the initially advantaged individuals. We are interested in computing the asymptotic value of this quantity. The following proposition gives the solution of System \eqref{eqdiff_pref}, after the change of variables.

\begin{proposition}\label{Solution_eqdiff_pref}
    The differential system (\ref{eqdiff_pref}) with initial condition $(b,b,0)$ for $b \in (0,1)$, has a unique solution $z_a \coloneqq (y_a(t),u_a(t),v_a(t))_{t\geq 0}$, written as follows:
\begin{equation}\label{eq-sol-syst-attract}
     \begin{cases}
     y_a(t) = \sigma^{-1} \left( \sigma(b)\exp\Big(\dfrac{s}{2}t\Big)\right) \\[10pt]
     \Delta_a(t) = g(y_a(t))/g(b)
     \\[15pt]
     u_a(t)+v_a(t)=b+\int_b^{y_a(t)}\frac{g(w)}{g(b)}dw,
    \end{cases}
\end{equation}
where $\sigma : y \mapsto \frac{y}{(1-y)^{1+s}}$ is a strictly increasing function from $(0,1)$ to $(0, \infty)$ and $g(y)=\left(\frac{1-y}{y}\right)^{\frac{2+s}{2s}}$ for any $y\in(0,1]$.
\end{proposition}

\begin{figure}[!ht]
\centering
\includegraphics[width=\textwidth]{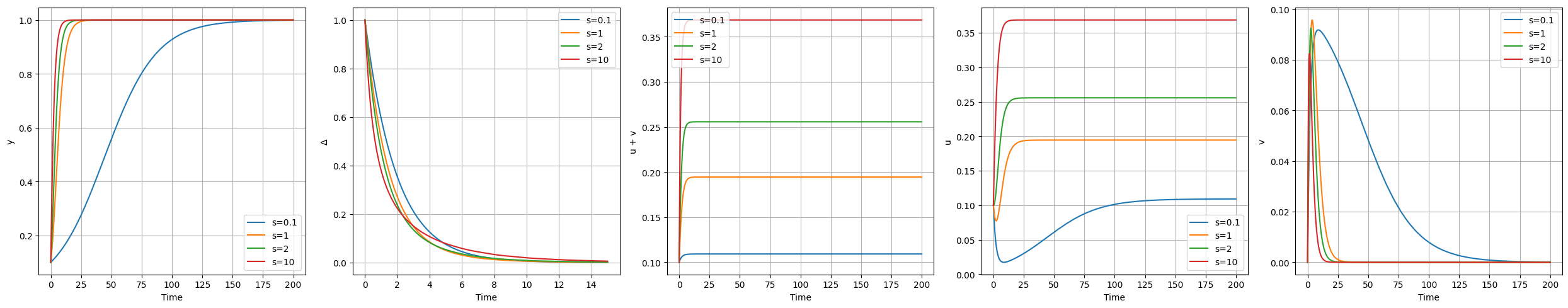}
\caption{Solutions in the increased attractiveness case of System (\ref{eqdiff_pref}) starting with $b = 0.1$ for different strengths of selection $s$. From left to right : trajectories of $y_a$, $\Delta_a$, $u_a+v_a$, $u_a$, $v_a$.}
\label{fig:attractive_s_comparison}
\end{figure}

    The function $t\rightarrow y_a(t)$ is always increasing, which corresponds to the fact that type $1$ induces an advantage over type $0$, so that in the end all individuals will be of type $1$. However one can notice that $u'_a(0)=(s-2)\frac{b(1-b)}{4(1+sb)}$. This shows that the case where $s=2$ is critical. In particular, whenever $s<2$, the weight of initially advantaged individuals among advantaged individuals initially decreases whereas it always increases whenever $s>2$ (this can be seen in the fourth graph of Figure \ref{fig:attractive_s_comparison}). This initial decline of $u_a$ occurs because, when selection is too weak, the decrease of $u_a$ from the death of advantaged individuals is not compensated in the beginning by the increase in $u_a$ from reproduction. This is due to the fact that, when the initial advantaged population is low, mothers are likely to choose a disadvantaged father as partner, and these fathers initially carry no genome coming from the initially advantaged individuals. Over time however, this genetic weight among disadvantaged individuals $v_a$ increases, which causes $u_a$ to increase as well. Besides, for any selection parameter $s>0$, the sum $u_a+v_a$ increases over time. These remarks are illustrated in Figure \ref{fig:attractive_s_comparison}.

\subsection{Homogamous choosiness}\label{sec-homo}

We now focus on the case of selection due to homogamous choosiness of one of the two parents.
The following theorem, proved in Appendix \ref{sec-homo-app}, is analogous to Theorem \ref{Convergence_pref}, but for the homogamous setting. Recall that in this scenario we set $s_0=0$ and $s_1=s>0$, so that type $1$ mothers preferentially choose a father of type $1$, while type $0$ mothers choose a father uniformly.
\begin{theorem} \label{Convergence_homo} Let $b \in (0,1)$. If the initial proportion of advantaged individuals $\frac{Y_0}{N}$ converges to $b\in (0,1)$  in probability when $N$ goes to infinity, then for any $c \in \mathbb{R}_{+}$, 
$$ \sup_{\, 0\leq t \leq c} \| Z_{\lfloor Nt \rfloor} - z_h(t) \| \longrightarrow_{N \rightarrow \infty} 0$$ in probability, where $z_h \coloneqq (y_h(t),u_h(t),v_h(t))_{t\geq 0}$ is the unique solution of the following differential equation: 
\begin{equation}\label{eqdiff_homo}
    \begin{cases}
      y' =  \dfrac{(s/2)y^2(1-y)}{1+sy} \\[10pt]
     u' = -\dfrac{u(1-y)}{1+sy} + \dfrac{1}{4}\biggl(\dfrac{2+sy}{1+sy}\biggl)(u(1-y)+vy) \\[10pt]
     v' =  -vy + \dfrac{1}{4}\biggl(\dfrac{2+sy}{1+sy}\biggl)(u(1-y)+vy),
    \end{cases}\,
\end{equation}
with initial condition $(b,b,0)$.
\end{theorem}

Analogously to the previous section, the subscript "$h$" now stands for "homogamous choosiness".
\begin{proposition}\label{Solution_eqdiff_homo}
The differential system \eqref{eqdiff_homo} with initial condition $(b,b,0)$ for $b \in (0,1)$, has a unique solution $z_h \coloneqq (y_h(t),u_h(t),v_h(t))_{t\geq 0}$, written as follows:
\begin{equation}\label{eq_sol_homo}
     \begin{cases}
     y_h(t) = \rho^{-1}\biggl(\rho(b)\exp(\frac{s}{2}t)\biggl) \\[10pt]
     \Delta_h(t) = g(y_h(t))/g(b)
     \\[15pt]
     u_h(t)+v_h(t)=b+\int_b^{y_h(t)}\frac{g(w)}{g(b)}dw,
    \end{cases}
\end{equation}
where $\rho: y \mapsto \exp(-1/y)(\frac{y}{1-y})^{(1+s)}$, is a strictly increasing function from $(0,1)$ to $(0,+\infty)$ and $g(y)=\left(\frac{1-y}{y}\right)^{\frac{2+s}{2s}}\exp\left(\frac{1}{sy}\right)$ for any $y\in(0,1]$.
\end{proposition}

\begin{figure}[!ht]
\centering
\includegraphics[width=\textwidth]{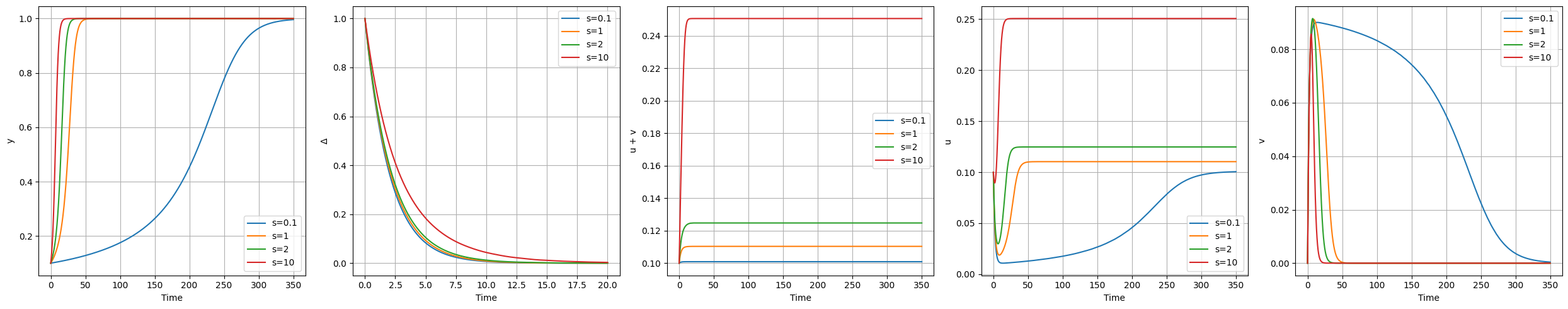}
\caption{Solutions of the homogamous case system (\ref{eqdiff_homo}) starting with $b = 0.1$ for different strengths of selection $s$. From left to right : trajectories of $y_h$, $\Delta_h$, $u_h+v_h$, $u_h$, $v_h$.}
\label{fig:homogamy_s_comparison}
\end{figure}

Figure \ref{fig:homogamy_s_comparison} shows the solutions of System \eqref{eqdiff_homo} as functions of time for several selection parameters $s>0$. As we observed for the increased attractiveness framework, for small values of $s$, the weight $u_h$ is initially decreasing with time. However in the case with homogamous choosiness, this decreasing phase depends both on the value of $s$ and on the initial value $y_0=b$, since $u'_h(0)=\frac{b(1-b)}{4(1+sb)}[sb-2]$. Finally, as mentioned for the increased attractiveness case, the quantity $u_h+v_h$ is always increasing.

\subsection{Viability selection}\label{sec-viability}
\begin{figure}[!ht]
\centering
\includegraphics[width=\textwidth]{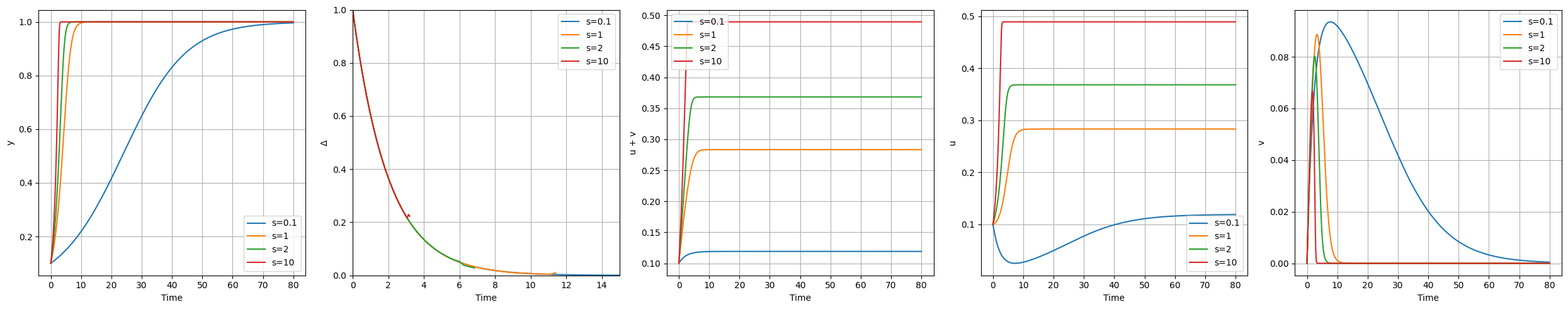}
\caption{Solutions of the viability selection case system (\ref{eqdiff_death}) starting with $b = 0.1$ for different strengths of selection $s$. From left to right: trajectories of $y$, $\Delta$, $u+v$, $u$, $v$.}
\label{fig:death_s_comparison}
\end{figure}

A particularity of the Moran model compared to the more classical Wright-Fisher model is that death of individuals is not a deterministic process, as shown in \cite{MuirheadWakeley2009} and in \cite{BrahaAguiar2026} in a biparental framework. Hence, selection can influence either the reproduction or the viability of individuals (or both) and the mathematical model makes a distinction between these phenomena. In particular, it enables efficiency comparisons between these different forms of selection in terms of transmission of genetic material. The biparental Moran model with viability selection has been introduced and studied in \cite{coronLeJan2025}. In this scenario, both parents are chosen uniformly randomly in the population and the advantage consists of a lower probability of being chosen to die. Namely, disadvantaged individuals die with a probability $(1+s)$ times higher than the advantaged ones. This leads, as shown in Proposition 2.7 of \cite{coronLeJan2025}, to the following limiting dynamics,  which is the equivalent of Systems \eqref{eqdiff_pref} and \eqref{eqdiff_homo} for the two reproductive selection patterns studied previously:

\begin{equation}\label{eqdiff_death}
    \begin{cases}
      y_d' = \dfrac{sy_d(1-y_d)}{y_d + (1+s)(1-y_d)} \\[10pt]
     u_d' = \dfrac{u_d}{2}+ \dfrac{u_d+v_d}{2}y_d - \dfrac{u_d}{y_d + (1+s)(1-y_d)}\\[10pt]
     v_d' = \dfrac{v_d}{2}+ \dfrac{u_d+v_d}{2}(1-y_d) - \dfrac{v_d(1+s)}{y_d + (1+s)(1-y_d)}
    \end{cases}
\end{equation}
with $(y_d(0),u_d(0),v_d(0))=(b,b,0)$.
In this viability selection framework we obtain that the quantity of genome transmitted by advantaged ancestors through time satisfies

\begin{equation}\label{eq_sol_death}
     \begin{cases}
     y_d(t) = \eta^{-1}\biggl(\eta(b)\exp(st)\biggl) \\[10pt]
     \Delta_d(t) = g(y_d(t))/g(b)
     \\[15pt]
     u_d(t)+v_d(t)=b+\int_b^{y_d(t)}\frac{g(w)}{g(b)}dw,
    \end{cases}
\end{equation}
where $\eta: y \mapsto \frac{y^{1+s}}{1-y}$, is a strictly increasing function from $(0,1)$ to $(0,+\infty)$ and $g(y)=\frac{(1-y)^{\frac{1}{2s}}}{y^{\frac{1+s}{2s}}}$ for any $y\in(0,1]$. Equation \eqref{eq_sol_death} can be obtained from  calculations given in the proof of Proposition 2.7 of \cite{coronLeJan2025}. Figure \ref{fig:death_s_comparison} shows the solutions of the System \eqref{eqdiff_death} as functions of time for several selection parameters $s>0$. 

\subsection{The infinite selection case}\label{sec:comparison_infinite_s}

In this section we will consider the case where selection is infinite ($s =\infty$). In the increased attractiveness case, this corresponds to a population where the mother, no matter her type, will always choose a partner in the advantaged class. For the choosiness case, homogamous (i.e. advantaged) mothers will always mate within their group whereas mothers of the other type will still choose their partner uniformly among the population. For the increased attractiveness case we get

\begin{proposition}\label{s_inf_pref}  When $s=\infty$, the System \eqref{eqdiff_pref} has a unique solution starting from $(b,b,0)$, given by
\begin{equation}\label{solution_eqdiff_pref_inf}
    \begin{cases}
      y_a(t) = 1 - (1-b)e^{-t/2} \\[10pt]
      u_a(t) = y_a(t) \, \sqrt{\dfrac{b}{1-b}} \biggl[\sqrt{\dfrac{1-y_a(t)}{y_a(t)}} + \arctan\biggl(\sqrt{\dfrac{1-b}{b}}\biggl) - \arctan\biggl(\sqrt{\dfrac{1-y_a(t)}{y_a(t)}}\biggl)\biggl] \\[10pt]
      v_a(t) = (1-y_a(t)) \sqrt{\dfrac{b}{1-b}} \biggl[ \arctan\biggl(\sqrt{\dfrac{1-b}{b}}\biggl) - \arctan\biggl(\sqrt{\dfrac{1-y_a(t)}{y_a(t)}}\biggl)\biggl].
    \end{cases}
\end{equation}
\end{proposition}

\begin{proof} First the System \eqref{eqdiff_pref} when $s$ is infinite gives that $y_a' = \dfrac{1}{2}(1-y_a)$, which gives that $(1-y_a)'=-\dfrac{1}{2}(1-y_a)$ and therefore that $1-y_a(t)=(1-b)\exp(-t/2)$ which gives the first line of Equation \eqref{solution_eqdiff_pref_inf}. Next from the second line of Equation \eqref{eq-sol-syst-attract}, $\Delta_a(t)=\sqrt{\frac{1-y_a(t)}{y_a(t)}}$ and from the third line of Equation \eqref{eq-sol-syst-attract}, if $s=\infty$, then \begin{align*}
  u_a(t)+v_a(t)&=b+\int_b^{y_a(t)}\sqrt{\frac{1-w}{w}}dw\\&=\sqrt{\dfrac{b}{1-b}} \biggl[\sqrt{y_a(t)(1-y_a(t))} - \arctan\biggl(\sqrt{\dfrac{1-y_a(t)}{y_a(t)}}\biggl)+ \arctan\biggl(\sqrt{\dfrac{1-b}{b}}\biggl)\biggl]
\end{align*} 
Using that $$u=y(1-y)\left[\Delta+\frac{u+v}{1-y}\right]$$ (this equality being true at all times and for any reproduction model) gives the second equation of System \eqref{solution_eqdiff_pref_inf} and using that $v=u+v-u$ gives the last equation of System \eqref{solution_eqdiff_pref_inf}.
\end{proof}

Analogously,

\begin{proposition}\label{s_inf_homo}
When $s=\infty$, the System \eqref{eqdiff_homo} has a unique solution starting from $(b,b,0)$, given by
\begin{equation}\label{solution_eqdiff_homo_inf}
    \begin{cases}
      y_h(t) = \dfrac{be^{t/2}}{1-b(1-e^{t/2})} \\[10pt]
      u_h(t) =  y_h(t)\sqrt{\dfrac{b}{1-b}}\left[\sqrt{\dfrac{1-y_h(t)}{y_h(t)}}+ \arctan\left(\sqrt{\dfrac{1-b}{b}}\right)-\arctan\left(\sqrt{\dfrac{1-y_h(t)}{y_h(t)}}\right)\right] \\[10pt]
      v_h(t)=
    (1-y_h(t))\sqrt{\dfrac{b}{1-b}}\left[\arctan\biggl(\sqrt{\dfrac{1-b}{b}}\biggl) - \arctan\biggl(\sqrt{\dfrac{1-y_h(t)}{y_h(t)}}\biggl)\right].
    \end{cases}
\end{equation}
\end{proposition}

\begin{proof} The first line of the System \eqref{eqdiff_homo} for $s=+\infty$ gives that $y_h' = \dfrac{1}{2}y_h(1-y_h)$ which gives that $y_h'\left[\dfrac{1}{y_h}+\dfrac{1}{1-y_h}\right] = \dfrac{1}{2}$ therefore $\dfrac{y_h(t)}{1-y_h(t)}=\dfrac{b}{1-b}\exp(t/2)$ which gives the first line of the System \eqref{solution_eqdiff_homo_inf}. The end of the proof is exactly the same as in Proposition \ref{s_inf_pref}, using the second and third equations of System \eqref{eqdiff_homo}.
\end{proof}

The solutions of Propositions \ref{s_inf_pref} and \ref{s_inf_homo}, verify that there exist some functions $\phi,\psi$ such that $u_i(t)=\phi(y_i(t))$ and $v_i(t)=\psi(y_i(t))$ for both $i=a,h$. So the genetic weights dynamics is the same in both models, up to an explicit time change linked to the speed of increase of the proportion of advantaged individuals.

The viability selection model presented in \cite{coronLeJan2025} can also be considered when selection is infinite. In this case, both parents are still chosen uniformly randomly, but the individual which dies is always chosen among the disadvantaged subpopulation. Proposition 2.7. in \cite{coronLeJan2025} gives the corresponding limiting differential system solution for $s>0$ and, if we let $s \rightarrow \infty$, we obtain the following expression:
\begin{equation}\label{solution_eqdiff_death_inf}
    \begin{cases}
      y_d(t) = be^t \textbf{1}_{t \leq -\log(b)} + \textbf{1}_{t > -\log(b)} \\[10pt]
     u_d(t) = b e^{t/2}(1+ be^t-be^{t/2})\textbf{1}_{t \leq -\log(b)} + (2\sqrt{b} - b)\textbf{1}_{t > -\log(b)}\\[10pt]
     v_d(t) = b(1-be^t)(e^{t/2} - 1)\textbf{1}_{t \leq -\log(b)}.
    \end{cases}
\end{equation}
As opposed to the previous cases, invasion of the advantaged individuals takes place here in finite time, $t = -\log(b)$.

\subsection{Comparison between different forms of selection}\label{sec-comparison}

In this subsection we compare the impact of the three forms of selection on genome transmission. 
In all cases
$$y(t) \underset{t\to\infty}{\to} 1 \qquad \text{ and} \qquad v(t)\underset{t\to\infty}{\to} 0, $$
which is consistent with the fact that, eventually, all individuals in the population will be advantaged. Similarly, Propositions \ref{Solution_eqdiff_pref}, \ref{Solution_eqdiff_homo}  and Equation \eqref{eq_sol_death} imply that the quantities $u_a(t)$, $u_h(t)$, $u_d(t)$ respectively converge, when $t$ goes to infinity, towards the constants
\begin{equation}\label{eq-limit-u}\begin{aligned}u_a(+\infty)&=b+\left(\frac{b}{1-b}\right)^{\frac{2+s}{2s}}\int_b^{1}\left(\frac{1-w}{w}\right)^{\frac{2+s}{2s}}dw, \quad \text{and}\\ u_h(+\infty)&=b+\left(\frac{b}{1-b}\right)^{\frac{2+s}{2s}}\int_b^{1} \left(\frac{1-w}{w}\right)^{\frac{2+s}{2s}}\frac{e^{1/sw}}{e^{1/sb}}dw, \quad \text{and}\\u_d(+\infty)&=b+\left(\frac{b}{1-b}\right)^{\frac{2+s}{2s}}\int_b^1\frac{(1-w)^{\frac{1}{2s}}}{w^{\frac{1+s}{2s}}}\frac{(1-b)^{\frac{1+s}{2s}}}{b^{\frac{1}{2s}}}dw.\end{aligned}\end{equation} 

To sum up, for all three models, the asymptotic genetic contribution of the initially
advantaged ancestors takes the unified form
\begin{equation}\label{eq:unified-u-infty}
u(\infty)\ =\ b+\left(\frac{b}{1-b}\right)^{\!\frac{2+s}{2s}}
\int_b^1 \left(\frac{1-w}{w}\right)^{\!\frac{2+s}{2s}}\!f(w)\,dw,
\end{equation}
where the function $f$ depends on the form of selection. 
More generally, for all three models, the dynamics of the genetic contribution of advantaged ancestors satisfies :
\begin{equation}\label{eq:unified-u+v}
u(t)+v(t)\ =\ b+\left(\frac{b}{1-b}\right)^{\!\frac{2+s}{2s}}
\int_b^{y(t)}\left(\frac{1-w}{w}\right)^{\!\frac{2+s}{2s}}f(w)dw.
\end{equation}
Table~\ref{tab:summary}
summarises the main quantities of the three models.

\begin{table}[!ht]
\centering
\small
\renewcommand{\arraystretch}{1.7}
\resizebox{\textwidth}{!}{%
\begin{tabular}{|l|c|c|c|}
\hline
 & \textbf{Attractiveness} & \textbf{Homogamous choosiness} & \textbf{Viability} \\
 & (Sec.~\ref{sec-attract}) & (Sec.~\ref{sec-homo}) & (Sec.~\ref{sec-viability}) \\
\hline
Selection rule
 & $s_0=s_1=s$
 & $s_0=0,\ s_1=s$
 & death rate $\times(1+s)$ if disadv. \\
\hline
$y'(t)$
 & $\dfrac{(s/2)\,y(1-y)}{1+sy}$
 & $\dfrac{(s/2)\,y^2(1-y)}{1+sy}$
 & $\dfrac{s\,y(1-y)}{1+s(1-y)}$ \\
\hline
$\phi$ in $y(t)=\phi^{-1}\!\bigl(\phi(b)\,e^{c_s t}\bigr)$
 & $\dfrac{y}{(1-y)^{1+s}}$, $c_s=\tfrac{s}{2}$
 & $e^{-1/y}\!\left(\dfrac{y}{1-y}\right)^{\!1+s}\!,\ c_s=\tfrac{s}{2}$
 & $\dfrac{y^{1+s}}{1-y}$, $c_s=s$ \\
\hline
$f(w)$ in \eqref{eq:unified-u-infty} and \eqref{eq:unified-u+v} 
 & $f_a(w)=1$
 & $f_h(w)=\exp\!\Bigl(\dfrac{1}{sw}-\dfrac{1}{sb}\Bigr)$
 & $f_d(w)=\dfrac{w^{1/(2s)}(1-b)^{(1+s)/(2s)}}{(1-w)^{(1+s)/(2s)}\,b^{1/(2s)}}$ \\
\hline
\multirow{2}{*}{$u(\infty)$ at $s=\infty$}
 & \multicolumn{2}{c|}{\multirow{2}{*}{$\displaystyle\sqrt{\dfrac{b}{1-b}}\,\arctan\!\sqrt{\dfrac{1-b}{b}}$}}
 & \multirow{2}{*}{$2\sqrt{b}-b$} \\
 & \multicolumn{2}{c|}{} &  \\
\hline
Time to reach $u(\infty)$ at $s=\infty$
 & \multicolumn{2}{c|}{$+\infty$}
 & finite ($t=-\log b$) \\
\hline
\end{tabular}}
\caption{Summary of the three selection models main properties.
}
\label{tab:summary}
\end{table}

The unified formula \eqref{eq:unified-u-infty} reveals that the three models share the same mathematical skeleton: the asymptotic genetic weight is determined by integrating a product of the genealogical distance function $\left(\frac{1-w}{w}\right)^{(2+s)/(2s)}$ and a model-specific weight $f(w)$ over the allele frequency interval $[b,1]$. These different fronts of invasion will lead to different limiting values of $u(t)$ when $t$ goes to infinity, which can be interpreted as the fact that the strength and form of selection will have an impact on the quantity of genome asymptotically transmitted by initially advantaged individuals.

\subsubsection{Trajectories comparison for same selection parameter}\label{sec:traj-comparison}

The unified formula \eqref{eq:unified-u-infty} as well as Table \ref{tab:summary} make it easy to compare the impacts of the different forms of selection. In particular we prove that for the same selection parameter $s$, selection due to increased attractiveness is at all times stronger than selection due to homogamous choosiness, in the sense that it leads to a larger quantity of transmitted genome and a larger number of advantaged individuals.

\begin{proposition}\label{prop-comparison} When the selection parameter $s$ and the initial population $b$ are the same in both models, then for all $t>0$
\begin{description}
\item[$(i)$] $y_h(t)<y_a(t)$
\item[$(ii)$] $u_h(t)+v_h(t)<u_a(t)+v_a(t)$.
\end{description}
\end{proposition}

\begin{proof} 
From Table \ref{tab:summary}, the dynamics of the proportion of advantaged individuals in the increased attractiveness case satisfies $y_a'=\frac{s}{2}\frac{y_a(1-y_a)}{1+sy_a}$, whereas in the homogamous choosiness case it satisfies $y_h'=\frac{s}{2}\frac{y_h^2(1-y_h)}{1+sy_h}$. Therefore if $y_a(t)=y_h(t)$ then $y'_a(t)>y'_h(t)$ which gives point $(i)$. 

Now using the formalism of Equation \eqref{eq:unified-u+v} and Table \ref{tab:summary}, since $y > b$ implies $f_h(y)<f_a(y)$ we get that for any $t>0$,
\begin{align*}
u_h(t)+v_h(t) &= \ b+\left(\frac{b}{1-b}\right)^{\!\frac{2+s}{2s}}
\int_b^{y_h(t)}\left(\frac{1-w}{w}\right)^{\!\frac{2+s}{2s}}f_h(w)dw \\
&< b+\left(\frac{b}{1-b}\right)^{\!\frac{2+s}{2s}}
\int_b^{y_h(t)}\left(\frac{1-w}{w}\right)^{\!\frac{2+s}{2s}}f_a(w)dw\\ &< b+\left(\frac{b}{1-b}\right)^{\!\frac{2+s}{2s}}
\int_b^{y_a(t)}\left(\frac{1-w}{w}\right)^{\!\frac{2+s}{2s}}f_a(w)dw \\&= u_a(t)+v_a(t),
\end{align*}
which proves $(ii)$.
\end{proof}

\begin{figure}
    \centering
    \includegraphics[width=\textwidth]{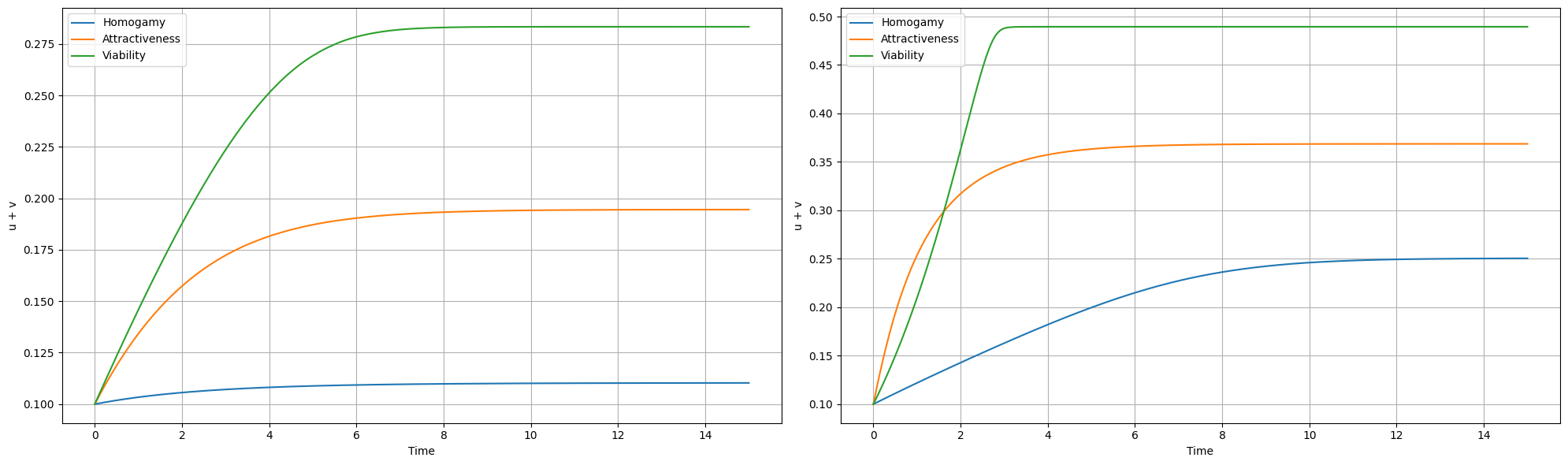}
    \caption{Trajectories of the genetic contribution of the initially advantaged ancestors ($u + v$), when $s=1$ (left) and $s=10$ (right), for the homogamous selection case (in blue), the increased attractiveness selection case (in orange) and the viability selection case (in green).}
    \label{fig:Comparison_s_fixed}
\end{figure}

Figure \ref{fig:Comparison_s_fixed} shows the dynamics of the genetic contribution of the initially advantaged ancestors for the three different types of selection, for different values of selection parameters. In particular, it illustrates the results stated in Proposition \ref{prop-comparison}: we see that for same selection strength $s$ the population size increases more rapidly when selection is materialised by an increased attractiveness, the genetic contribution of the initially advantaged individuals is larger at all times, and the asymptotic contribution is also larger. This is due to the fact that under increased attractiveness each mating event is biased in favour of the advantaged type, whereas under homogamous choosiness only homogamous mothers generate reproduction events biased towards their own (advantaged) type. 

System \eqref{eq_sol_death} does not allow for a simple comparison of $u_d(t)+v_d(t)$ with the previous quantities $u_a(t)+v_a(t)$ and $u_h(t)+v_h(t)$ at any time $t$. Similarly, the comparison of the number of advantaged individuals between the different forms of selection is not as simple as for the comparison of the two forms of reproductive selection. 
In particular, for some values of selection parameter $s$ and initial advantaged proportion $b$, the growth of the advantaged population in the viability selection model will be slower at the beginning as we can see for instance for $s=10$ and $b=0.1$ in Figure \ref{fig:Comparison_s_fixed} (right). This is due to the fact that in the viability selection model, if the initial advantaged population is too small, most of the parents will be disadvantaged and therefore their children too, whereas in the increased attractiveness setting, one of the parents will be chosen among the advantaged class with high probability (if $s$ is large enough). However, for any initial condition $b \in (0,1)$, we can check that $y_h(t) \leq y_d(t)$ for every $t \geq 0$ using the same argument as for point $(i)$ of Proposition \ref{prop-comparison}.

\subsubsection{Comparison of asymptotic advantaged ancestors contributions}\label{sec:asympt-comparison}

\paragraph{The finite selection case}

\begin{proposition}\label{prop:same_s_limit}
When the selection parameter is finite and the same in all models, $u_h(+\infty) < u_a(+\infty)<u_d(+\infty)$.
\end{proposition}

\begin{proof}
The strict pointwise inequality $f_h(y) < f_a(y)<f_d(y)$ on $(b,1)$ yields $u_h(+\infty) < u_a(+\infty) < u_d(+\infty)$ as in the proof of Proposition \ref{prop-comparison}, but taking $y(t)=1$ in both models.
\end{proof}

\begin{figure}[!ht]
  \centering
  \includegraphics[width=0.5\textwidth]{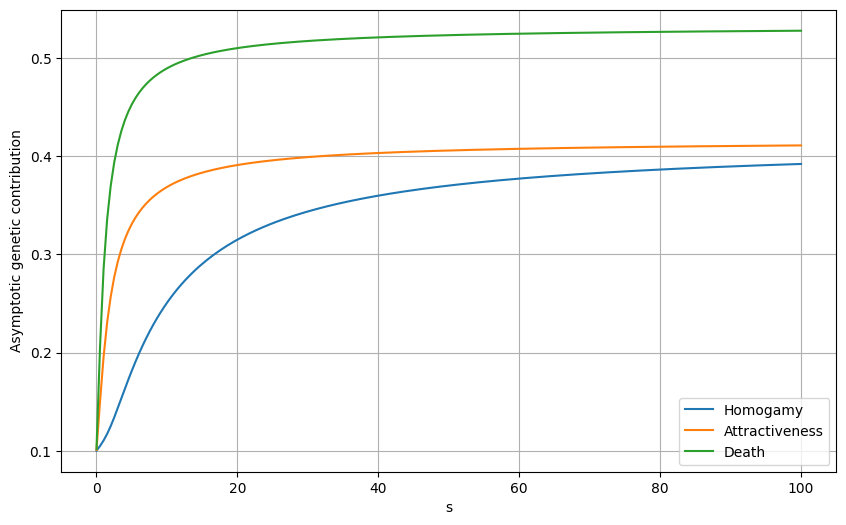}
  \caption{$u_a(\infty)$, $u_h(\infty)$ and $u_d(\infty)$ for $b=0.1$ as functions of $s$.}
\label{fig:u_inf_inf_depending_on_s}
\end{figure}

\begin{figure}
    \centering
    \includegraphics[width=\textwidth]{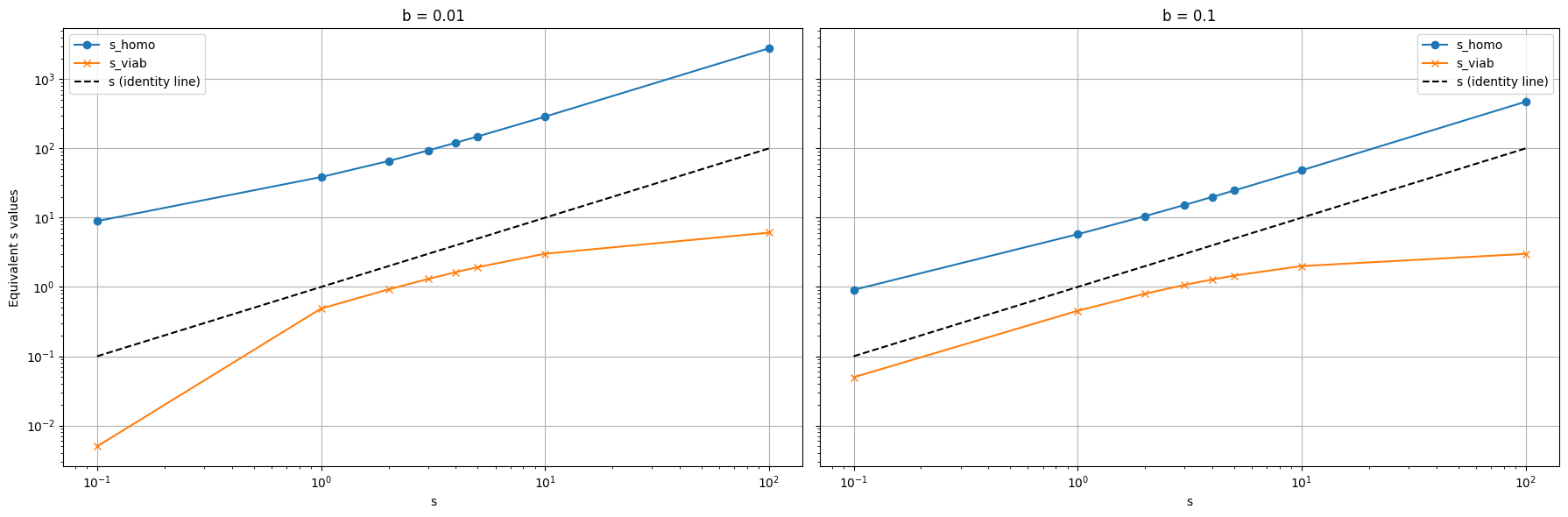}
    \caption{Equivalent selection parameters $s_{\mathrm{homo}}$ and $s_{\mathrm{viab}}$ for $b=0.01$ (left) and $b=0.1$ (right).}
    \label{fig:s_eq}
\end{figure}

These results prove that selection due to increased attractiveness is more efficient than selection due to homogamous choosiness. Equation \eqref{eq-limit-u} also allows to define an equivalent selection parameter for the homogamous selection model, that would lead to the same asymptotic contribution of initially advantaged individuals than the increased attractiveness model with selection parameter $s$. From the previous results, we expect this equivalent selection parameter to be larger than $s$. This notion of equivalent selection parameter is presented in Figure \ref{fig:s_eq} of the Discussion (Section \ref{sec:discussion}). Proposition \ref{prop:same_s_limit} also indicates that viability selection has a stronger asymptotic impact than increased attractiveness or homogamous choosiness selection on genetic transmission, for the same value of selection parameter. These results are illustrated in Figure \ref{fig:u_inf_inf_depending_on_s} for several values of $s$ and $b=0.1$.

\paragraph{The infinite selection case}

If we now let $t \rightarrow \infty$ in the solutions obtained in Propositions \ref{s_inf_pref} and \ref{s_inf_homo} we get the following
\begin{corollary}\label{cor-s-infty}
In the infinite selection case where $s=\infty$,
\begin{equation*}
        \lim_{t\rightarrow\infty}u_a(t) = \lim_{t\rightarrow\infty}u_h(t)=\sqrt{\frac{b}{1-b}}\arctan\biggl(\sqrt{\dfrac{1-b}{b}}\biggr). 
    \end{equation*}
\end{corollary}

\begin{proof}
As both $y_a(t)$ and $y_h(t)$ converge to $1$ when $t$ goes to infinity, the second line respectively in Equations \eqref{solution_eqdiff_pref_inf} and \eqref{solution_eqdiff_homo_inf} give the result.
\end{proof}

We can follow suit with the death model in Equation \eqref{solution_eqdiff_death_inf} to get
\begin{equation}\label{eq_u_select_death}
    u_d(t) \underset{t\to\infty}{\longrightarrow} 2\sqrt{b} - b.
\end{equation}

\begin{figure}[!ht]
  \centering
  \includegraphics[width=\textwidth]{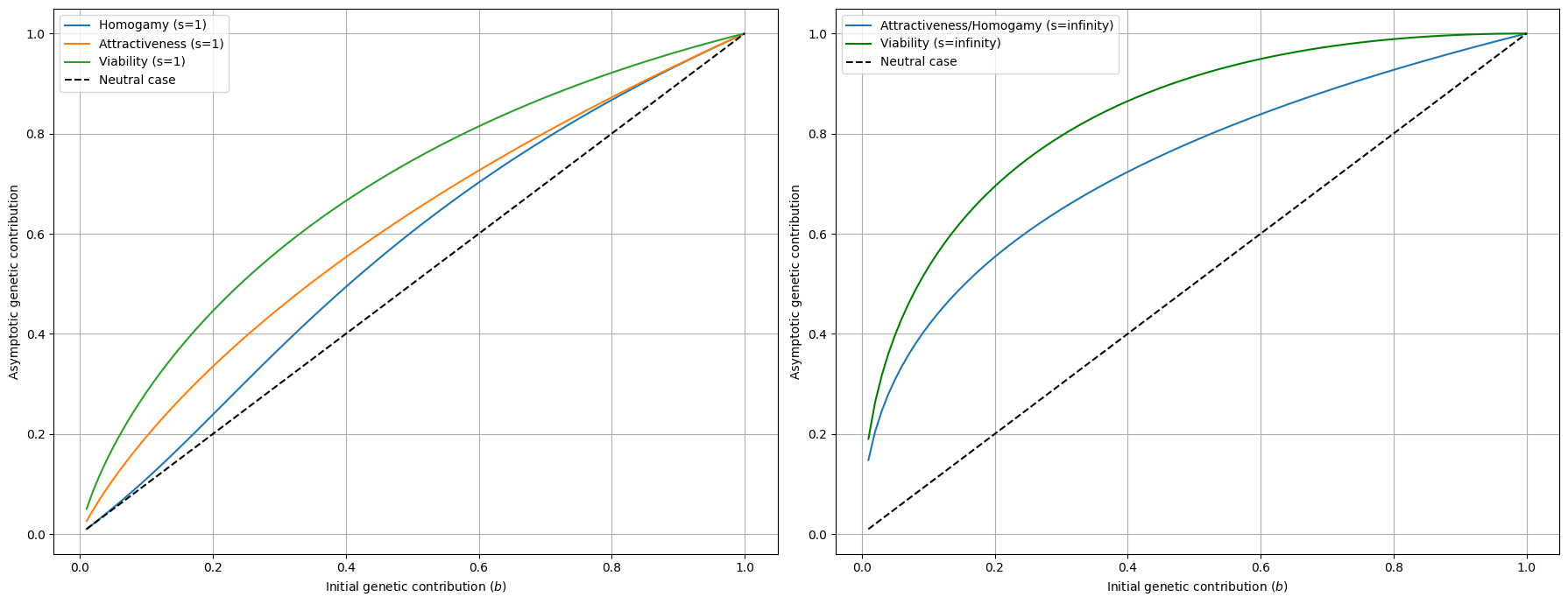}
  \caption{$u_a(\infty)$, $u_h(\infty)$ and $u_d(\infty)$ for $s=1$ (left) and $s=\infty$ (right) as a function of $b$.}
\label{fig:u_inf_inf_depending_on_b}
\end{figure}

These quantities give the maximal value of the asymptotic genetic contribution of an initially advantaged population, as a function of its initial proportion of the population. Both the increased attractiveness and homogamous choosiness settings converge to the same limit weight, which is smaller than the value in \eqref{eq_u_select_death} for all $b \in (0,1)$. For instance, if we start with $b = 1\%$ advantaged individuals, the asymptotic weight in the viability selection case will be of $19\%$ as opposed to the $14.8\%$ in the other two cases. Furthermore, note that the limit in \eqref{eq_u_select_death} is attained in finite time, contrarily to the other two. We can observe this dominance in \ref{fig:u_inf_inf_depending_on_b} (right), where we have represented this weight as a function of $b$, for a finite and infinite $s$.

\paragraph{Equivalent selection parameters.}
Our results enable the definition of equivalent selection parameters. Let us assume that a first population starts with a proportion $b=1\%$ of advantaged individuals and that this advantage is due to an increased attractiveness, with parameter $s$, as described in this article. Now let us consider two other populations, also starting with a proportion $b=1\%$ of advantaged individuals, but for which the advantage is respectively due to homogamous choosiness and to increased viability, with respective parameters $s_{\mathrm{homo}}$ and $s_{\mathrm{viab}}$. Then using Equation \eqref{eq-limit-u}, the parameters $s_{\mathrm{homo}}$ and $s_{\mathrm{viab}}$ leading to the same asymptotic transmission of genome can be computed. However these equivalent selection parameters depend on the initial proportion $b$ of advantaged individuals. As expected from our obtained result, we have $s_{homo}>s>s_{viab}$. They are plotted as functions of $s$ in Figure \ref{fig:s_eq}, for two different values of $b$.

\section{Discussion}\label{sec:discussion}

In this work, we examined how the form of selection impacts genome transmission during a selective sweep. Building on a previous approach to viability selection, we modified the framework introduced by \cite{CoronLeJan2022JMB} to account for two forms of reproductive selection: increased attractiveness and homogamous choosiness. Three main results emerged from this work. First, our framework enables an explicit prediction of the asymptotic genetic contribution of initially advantaged individuals. Second, we show that the selective sweep is influenced by the form of selection, and in particular that in terms of genome transmission the advantage conferred by a selected allele is stronger for viability selection, followed by attractiveness, with homogamous choosiness resulting in the lowest genetic contribution after fixation of the beneficial allele. Third, we find that when the selection parameter goes to infinity, the differences in the efficiency of genome transmission between attractiveness and homogamous choosiness vanish.

This ordering can be interpreted intuitively. Viability selection affects life-time reproductive success, as favoured individuals live longer have more opportunities to reproduce, regardless of the mating partner involved. This accounts for the systematically higher asymptotic genetic contribution for this form of selection. Increased attractiveness biases every mating event in the population in favour of the advantaged type, whereas homogamous choosiness operates only when choosy individuals are selected to mate. The form of the mating preference matters most in the regime of moderate selection, and the two preferences become indistinguishable in the limit of strong selection. 

Our predictions relate directly to recent observations of genome-wide diversity patterns following selective events, even in regions that are not linked to the locus under selection \citep{Ollivieretal2025}. We predict that the magnitude of such reductions will also depend on whether selection acts primarily through viability or through mating preferences. For hermaphroditic taxa, where sexual selection can operate on both male and female function simultaneously (\cite{Leonard2006,Beekmanetal2016}), the differential asymptotic contributions we quantify offer testable hypotheses. Specifically, lineages under strong viability-based selection should show more pronounced reductions in neutral diversity than those where similar allele-frequency changes arise through attractiveness or homogamous choosiness. Controlled introgression experiments, such as those tracking admixed genomes across generations in laboratory populations of hermaphroditic organisms (e.g., Caenorhabditis nematodes, freshwater snails, or self-incompatible plants), could directly measure these differences and validate the predicted ranking of genetic weights across selection modes.

Most population genetics tools have been developed using the Wright-Fisher model, and the Moran model has been shown to be equivalent to the Wright-Fisher model up to an appropriate time change in uniparental populations. In biparental models, however, as shown in \cite{CoronLeJan2022JMB}, these two classical models are no longer equivalent, owing to a difference between the (supercritical) laws of the number of offspring per individual in the two frameworks. The advantage of the Moran model is that it more readily allows specific processes of selection to be defined explicitly. In the uniparental Moran model without mutation, selection on death or on reproduction leads to similar rescaled dynamics of the allelic frequencies at the locus under selection (\cite{MuirheadWakeley2009}). Asymptotic allele frequencies under mutation-selection equilibrium in the biparental Moran model have also been characterised in \cite{BrahaAguiar2026}, and show distinct (though not very different) behaviours between selection on reproduction and viability. These works, however, focus on the dynamics at the loci under selection, and not on genome-wide consequences at effectively neutral loci. Whereas these works characterise allele-frequency dynamics at the selected locus, our analysis quantifies the consequences of the mode of selection on patterns of ancestral transmission. 

The present work opens several directions for future research. First, the two mating preferences studied here are particular cases of a more general framework in which each genotypic class has its own preference strength: advantaged individuals exhibit a preference parameter $s_1$ and disadvantaged ones a parameter $s_0$. Analysing arbitrary combinations ($s_0$, $s_1$) would yield a unified treatment of both models and reveal how the relative preference intensities of the two classes jointly shape genome transmission. Second, we restricted our analysis to hermaphroditic populations; extending the model to populations with separate sexes would require specifying preference parameters for each sex and would therefore lead to a richer parameter space, while being closer to many natural systems in which mate choice is predominantly female. Third, our results are derived under the assumption of a positive initial proportion b of advantaged individuals; a natural complement would be to condition on the allele starting as a single copy, and to characterise the genetic contribution of that unique ancestor conditional on eventual fixation, which is the setting most directly relevant to the study of selective sweeps. This question is the object of the work in progress \cite{CoronHenard2026}.

Finally, our study could serve as the basis for statistical inference. Controlled introgression programmes, in which a wild or improved strain is introduced into a resident population, replicated across independent parcels with various and known initial frequencies, can provide a particularly relevant context for such an inference. The observation of the asymptotic proportion of introgressed genome would indeed enable simultaneous inference of both the strength and the form of selection. Furthermore, when the practical objective is to maximise introgression (for instance to maximise yield of cultivated varieties), the notion of equivalent selection parameters illustrated in Figure \ref{fig:s_eq} provides a quantitative criterion for comparing the impacts of hybridisation strategies leading to different forms of selection.

\section*{Acknowledgements}
J.M-S. acknowledges the support of the grant ``IDEX Bourse Internationale de Master 2024-2025" of the Université Paris-Saclay for his master studies during which this project was started.

\appendix

\section{Proofs}

\subsection{General dynamics}\label{sec-Dyn-Z-app}

The following proposition gives the dynamics of the stochastic process $(Z_n)_{n\in \mathbb{N}}$.
\begin{proposition}\label{prop:dynamics_Z}
The sequence $(Z_n)_{n \in \mathbb{N}}$ satisfies the following recursion property:
\begin{equation}\label{eq-dynamics}
    \begin{aligned}
Z_{n+1} = Z_n + \frac{1}{N} &\biggl[ \biggl(0, - \sum_{l' \in \mathcal{Y}_0} W_n(\kappa_n, l') + \sum_{l' \in \mathcal{Y}_0} \frac{W_n(\mu_n, l') + W_n(\pi_n, l')}{2}, 0 \biggr) \mathbf{1}_{\nu_n\in \mathcal{Y}_n, \kappa_n \in \mathcal{Y}_n} \\
&+ \biggl( 0, 0, - \sum_{l' \in \mathcal{Y}_0} W_n(\kappa_n, l') + \sum_{l' \in \mathcal{Y}_0} \frac{W_n(\mu_n, l') + W_n(\pi_n, l')}{2} \biggr) \mathbf{1}_{\nu_n\notin \mathcal{Y}_n, \kappa_n \not\in \mathcal{Y}_n} \\
&+ \biggl(1, \sum_{l' \in \mathcal{Y}_0} \frac{W_n(\mu_n, l') + W_n(\pi_n, l')}{2}, - \sum_{l' \in \mathcal{Y}_0} W_n(\kappa_n, l') \biggr) \mathbf{1}_{\nu_n\in \mathcal{Y}_n, \kappa_n \not\in \mathcal{Y}_n} \\
&+ \biggl(-1, - \sum_{l' \in \mathcal{Y}_0} W_n(\kappa_n, l'), \sum_{l' \in \mathcal{Y}_0} \frac{W_n(\mu_n, l') + W_n(\pi_n, l')}{2} \biggr) \mathbf{1}_{\nu_n\notin \mathcal{Y}_n, \kappa_n \in \mathcal{Y}_n} \biggr].
    \end{aligned}
\end{equation}
\end{proposition}

\begin{proof} This proposition results from the model description in Section \ref{sec:model}. Indeed, at each time step, one can distinguish four possibilities for the dynamics, depending on the type of the parent that transmits its type (i.e. the type of the new-born individual), and on the type of the dying individual. This corresponds to the four lines of Equation \eqref{eq-dynamics}. The first line corresponds to the case where both the parent that transmit its type and the dying individual are advantaged. In this case the number of advantaged individuals and the weight of initially advantaged individuals among disadvantaged individuals remain unchanged (as disadvantaged individuals do not participate to the birth-and-death event at this time step), and the weight of initially advantaged individuals among advantaged individuals is decreased by the weight of initially advantaged individuals in the dying individual, and increased by the mean of the weights of initially advantaged individuals among the two parents. The three other lines are obtained similarly.
\end{proof}

\subsection{Increased attractiveness}\label{sec-attract-app}

In this section we focus on the case where the two types differ in terms of attraction: at each time step an individual with type $1$ has a probability that is $1+s$ times higher to be chosen as father than an individual with type $0$, no matter what the mother's type is. Thus, the choice of the parents is done independently in this setting, which simplifies the analysis of the dynamics. For simplicity in this section we will drop the subscript "$a$" present in Section \ref{sec-attract}. The following proposition gives the transition probabilities of the number of advantaged individuals, $(Y_n)_{n\in\mathbb{Z}_+}$.

\begin{proposition}\label{prop_Y_Attract}
The stochastic process $(Y_n)_{n\in\mathbb{Z}_+}$ is a Markov chain such that $Y_n\in\{0,...,N\}$ for all $n\in\mathbb{Z_+}$. Furthermore,
\begin{itemize}
\item[$(i)$] For any $k\in\{0,...,N\}$, $\mathbb{P}(Y_{n+1}\in\{k-1,k,k+1\}|Y_n=k)=1$, and the transition probabilities of $(Y_n)_{n\in\mathbb{Z}_+}$ are given by:
     $$\mathbb{P}[Y_{n+1} = k+1 | Y_n = k] = \dfrac{1}{2}\dfrac{k}{N}\biggl(1-\dfrac{k}{N}\biggr)\biggl[1+\dfrac{1+s}{1+s(k/N)}\biggr]$$ and
     $$\mathbb{P}[Y_{n+1} = k-1 | Y_n = k] = \dfrac{1}{2}\dfrac{k}{N}\biggl(1-\dfrac{k}{N}\biggr)\biggl[1+\dfrac{1}{1+s(k/N)}\biggr].$$
    In particular this Markov chain is absorbed at $0$ and $N$.
\item[$(ii)$] Let $b\in(0,1)$. When $N$ goes to infinity, $$\mathbb{P}(T_N<+\infty|Y_0=\lfloor bN\rfloor)\rightarrow 1.$$ 
\end{itemize}
\end{proposition}

\begin{proof}
From Proposition \ref{prop:dynamics_Z} (or the model description made in Section \ref{sec:model}), $Y_{n+1}=Y_n+1$ if and only if the replaced individuals $\kappa_n$ is chosen in $I\setminus\mathcal{Y}_n$ and the individual that transmits its type is chosen in $\mathcal{Y}_n$. This gives that \begin{align*}\mathbb{P}(Y_{n+1}=Y_n+1|Y_n)&=\left(1-\frac{Y_n}{N}\right)\frac{1}{2}\left[\frac{Y_n}{N}+\frac{(1+s)Y_n}{N-Y_n+(1+s)Y_n}\right]\\&=\frac{1}{2}\left(1-\frac{Y_n}{N}\right)\frac{Y_n}{N}\left[1+\frac{1+s}{1+s\frac{Y_n}{N}}\right].\end{align*}
The two terms in the right-hand side of the first line of the previous equation respectively correspond to the case where the mother is advantaged and transmits its type (i.e. $\nu_n=\mu_n\in\mathcal{Y}_n$), and the case where the father is advantaged and transmits its type (i.e. $\nu_n=\pi_n\in\mathcal{Y}_n$). As the choice of the father is independent of the mother's type there is no need to take this latter type into account in this second case.
Similarly,
\begin{align*}\mathbb{P}(Y_{n+1}=Y_n-1|Y_n)&=\frac{Y_n}{N}\frac{1}{2}\left[\frac{N-Y_n}{N}+\frac{N-Y_n}{N-Y_n+(1+s)Y_n}\right]\\&=\frac{1}{2}\left(1-\frac{Y_n}{N}\right)\frac{Y_n}{N}\left[1+\frac{1}{1+s\frac{Y_n}{N}}\right].\end{align*}
Therefore 
\begin{align*}\mathbb{P}(Y_{n+1}=Y_n+1|Y_{n+1}\neq Y_n,Y_n)&=\frac{2+s\left(1+\frac{Y_n}{N}\right)}{4+s\left(1+\frac{2Y_n}{N}\right)}=f\left(\frac{Y_n}{N}\right),\end{align*} where $f(y)=\frac{2+s(1+y)}{4+s(1+2y)}$ for all $y\in[0,1]$. The function $f$ is decreasing on $[0,1]$ therefore \begin{align*}\mathbb{P}(Y_{n+1}=Y_n+1|Y_{n+1}\neq Y_n,Y_n)\geq f(0)=\frac{1}{2}+\frac{1}{2}\frac{s}{4+s}.\end{align*}
This implies that the skeleton of the Markov chain $(Y_n)_{n\in\mathbb{Z}_+}$ (i.e. the sequence of the values taken by this process when it changes state) stochastically dominates a biased random walk on $\{0,...,N\}$ that is stopped in $0$ and $N$ and jumps at each time step to the right with probability $\frac{1}{2}+\frac{1}{2}\frac{s}{4+3s}$ and to the left with probability $\frac{1}{2}-\frac{1}{2}\frac{s}{4+3s}$. This gives point $(ii)$ as the result is true for this biased and stopped random walk (from \cite{Feller}, Chapter 14).
\end{proof}

\begin{proof}[Proof of Theorem \ref{Convergence_pref}]
From Equation \eqref{eq-dynamics}, for all $n \geq 0 $, one has
\begin{equation*} 
\begin{aligned} 
\mathbb{E}[Y_{n+1}-Y_n|\mathcal{F}_n^Z]&=\mathbb{P}\left[Y_{n+1}=Y_n+1|\mathcal{F}_n^Z\right]-\mathbb{P}\left[Y_{n+1}=Y_n-1|\mathcal{F}_n^Z\right]\\& = \frac{1}{2}\frac{Y_n}{N}\frac{N-Y_n}{N}\frac{s}{1+s\frac{Y_n}{N}}
\end{aligned}
\end{equation*}

from Proposition \ref{prop_Y_Attract}.
Now, from Equation \eqref{eq-dynamics} again,

\begin{equation*}
\begin{aligned} 
\mathbb{E}[U_{n+1}-U_n|\mathcal{F}_n^Z]&=\mathbb{E}\left[-\!\!\sum_{l'\in\mathcal{Y}_0}W_n(\kappa_n,l')\mathbf{1}_{\kappa_n\in\mathcal{Y}_n}+\sum_{l'\in\mathcal{Y}_0}\frac{W_n(\mu_n,l')+W_n(\pi_n,l')}{2}\mathbf{1}_{\nu_n\in\mathcal{Y}_n}\Big|\mathcal{F}_n^Z\right]\\& = -\frac{Y_n}{N}\frac{U_n}{Y_n} +\mathbb{E}\left[\mathbf{1}_{\nu_n=\mu_n}\mathbf{1}_{\mu_n\in\mathcal{Y}_n}\sum_{l'\in\mathcal{Y}_0}\frac{W_n(\mu_n,l')+W_n(\pi_n,l')}{2}\right.\\&\phantom{-\frac{Y_n}{N}\frac{U_n}{Y_n} +\mathbb{E}[[[[}\quad\left.+\mathbf{1}_{\nu_n=\pi_n}\mathbf{1}_{\pi_n\in\mathcal{Y}_n}\sum_{l'\in\mathcal{Y}_0}\frac{W_n(\mu_n,l')+W_n(\pi_n,l')}{2}\right]\\&=-\frac{U_n}{N}+\frac{1}{4}\frac{Y_n}{N}\left[\frac{U_n}{Y_n}+\frac{(1+s)Y_n}{(1+s)Y_n+N-Y_n}\frac{U_n}{Y_n}\right.\\&\phantom{=-\frac{U_n}{N}+\frac{1}{4}\frac{Y_n}{N}[[[[[[[}\left.+\frac{N-Y_n}{(1+s)Y_n+N-Y_n}\frac{V_n}{N-Y_n}\right]\\&\phantom{=-\frac{U_n}{N}}+\frac{1}{4}\frac{(1+s)Y_n}{(1+s)Y_n+N-Y_n}\left[\frac{U_n}{Y_n}+\frac{U_n+V_n}{N}\right].
\end{aligned}
\end{equation*}

Similarly,
\begin{align*}
\mathbb{E}[V_{n+1}-V_n|\mathcal{F}^Z_n]&=\mathbb{E}\left[-\sum_{l'\in\mathcal{Y}_0}W_n(\kappa_n,l')\mathbf{1}_{\kappa_n\notin\mathcal{Y}_n}+\sum_{l'\in\mathcal{Y}_0}\frac{W_n(\mu_n,l')+W_n(\pi_n,l')}{2}\mathbf{1}_{\nu_n\notin\mathcal{Y}_n}\Big|\mathcal{F}_n^Z\right]\\& = -\frac{N-Y_n}{N}\frac{V_n}{N-Y_n} +\mathbb{E}\left[\mathbf{1}_{\nu_n=\mu_n}\mathbf{1}_{\mu_n\notin\mathcal{Y}_n}\sum_{l'\in\mathcal{Y}_0}\frac{W_n(\mu_n,l')+W_n(\pi_n,l')}{2}\right.\\&\phantom{-\frac{N-Y_n}{N}\frac{V_n}{N-Y_n} +\mathbb{E}}\quad\left.+\mathbf{1}_{\nu_n=\pi_n}\mathbf{1}_{\pi_n\notin\mathcal{Y}_n}\sum_{l'\in\mathcal{Y}_0}\frac{W_n(\mu_n,l')+W_n(\pi_n,l')}{2}\right]\\&=-\frac{V_n}{N}+\frac{1}{4}\frac{N-Y_n}{N}\left[\frac{V_n}{N-Y_n}+\frac{(1+s)Y_n}{(1+s)Y_n+N-Y_n}\frac{U_n}{Y_n}\right.\\&\phantom{-\frac{V_n}{N}+\frac{1}{4}\frac{N-Y_n}{N}[\frac{V_n}{N-Y_n}}\left.+\frac{N-Y_n}{(1+s)Y_n+N-Y_n}\frac{V_n}{N-Y_n}\right]\\&\phantom{=-\frac{V_n}{N}}+\frac{1}{4}\frac{N-Y_n}{(1+s)Y_n+N-Y_n}\left[\frac{V_n}{N-Y_n}+\frac{U_n+V_n}{N}\right].
\end{align*}

Therefore recalling that $Z_n=\frac{1}{N}(Y_n,U_n,V_n)$, we get:
$$\mathbb{E}(Z_{n+1}-Z_n|\mathcal{F}_n^Z)=\frac{1}{N}g(Z_n),$$
where \begin{align*}g(y,u,v)&=\Big(\frac{1}{2}\frac{sy(1-y)}{1+sy},\\&\quad\quad-\frac{u(1-y)}{1+sy}+ \frac{2+s}{4}\frac{vy + (1-y)u}{1+sy},\\&\quad\quad-\frac{vy(1+s)}{1+sy} + \frac{2+s}{4}\frac{vy+ (1-y)u}{1+sy}\Big).\end{align*}

The rest of the proof follows the steps of the proof of the analogous convergence result in \cite{coronLeJan2025} (Theorem 2.9).
More precisely, we use Doob's decomposition of the process $(Z_n)_{n\in\mathbb{Z}_+}$:  
$$ Z_n = Z_0 + \sum_{k=1}^nA_k + \frac{1}{N}\sum_{k=0}^{n-1} g(Z_k),$$
where $A_n = Z_n-\mathbb{E}(Z_{n}|\mathcal{F}_{n-1}^Z)$. The three components of the vector $M_n = \sum_{k=1}^n A_k$ are square-integrable martingales. Therefore from Doob's maximal inequality we obtain
\begin{equation*}
    \mathbb{E}\left(\sup_{0\leq n \leq \lfloor cN \rfloor} || M_n ||^2 \right) \leq 4\, \mathbb{E}\left( \,|| M_{\lfloor cN \rfloor}||^2 \right),
\end{equation*}
for all $c > 0$. We can easily prove that in between two consecutive time steps each coordinate of $(Y_n,U_n,V_n)$ varies at most one unit, so 
$$ || Z_n - Z_{n-1} || \leq \sqrt{3}/N, \qquad \text{for all} \; n \geq 1. $$ Using the previous remark and the orthogonality of the increments of a $L^2$-martingale we get
\begin{equation*}
    \mathbb{E}( || M_{\lfloor cN \rfloor}||^2 ) = \mathbb{E} ( ||\sum_{k=0}^{\lfloor cN\rfloor}A_k||^2 )= \mathbb{E}( \sum_{k=0}^{\lfloor cN\rfloor}||A_k||^2 ) \leq \frac{12c}{N}.
\end{equation*}
Hence, as $N \rightarrow \infty$
$$ \sup_{0 \leq n \leq \lfloor cN \rfloor} || M_n || = \sup_{0 \leq t \leq  c }  \left\lVert \sum_{k=1}^{\lfloor Nt \rfloor} A_k \right\rVert \longrightarrow 0 \quad \text{in probability}.$$
By definition of $z_t$, 
$$ z_t = (b,b,0) + \int_0^t g(z_s)ds, \quad \text{for all}\; \; t \geq 0. $$
Thus,
$$Z_{\lfloor Nt \rfloor} - z_t = (Y_0/N,Y_0/N,0) - (b,b,0) + \sum_{k=0}^{\lfloor Nt \rfloor} A_k + \int_0^t ( g(Z_{\lfloor Ns \rfloor}) - g(z_s) ) ds +R_N(t), $$  where $|R_N(t)|\leq\frac{\|g\|_{\infty}}{N}$.
Since the Jacobian of $g$ is bounded for $(y,u,v) \in [0,1]^3$, there exists a constant $K$ such that
$$|| Z_{\lfloor Nt \rfloor} - z_t || \leq || (Y_0/N,Y_0/N,0) - (b,b,0) || + \left\lVert \sum_{k=0}^{\lfloor Nt \rfloor} A_k \right\rVert + \int_0^t K \; || Z_{\lfloor Ns \rfloor} - z_s || ds + \frac{\|g\|_{\infty}}{N}.$$
From Gronwall's lemma, for all $t \in [0,c]$,
$$|| Z_{\lfloor Nt \rfloor} - z_t || \leq \left(|| (Y_0/N,Y_0/N,0) - (b,b,0) || + \left\lVert \sum_{k=0}^{\lfloor Nt \rfloor} A_k\right\rVert+ \frac{\|g\|_{\infty}}{N}\right) \; e^{Kc},$$
which gives the result.
\end{proof}

\begin{proof}[Proof of Proposition \ref{Solution_eqdiff_pref}] From Picard-Lindel\"{o}f (or Cauchy-Lipschitz) Theorem, the System \eqref{eqdiff_pref} admits a unique solution starting from $(b,b,0)$, denoted by $(z_t)_{t\geq0}=(y_t,u_t,v_t)_{t\geq0}$. We will now prove that this solution satisfies \eqref{eq-sol-syst-attract}. We first focus on the first equation of the differential System \eqref{eqdiff_pref}, which is autonomous. Since the right-hand side of this first equation is equal to $0$ when $y\in\{0,1\}$, we have that if $b \in (0, 1)$, then $y_t \in (0, 1)$ for all $t\geq 0$. Therefore the first equation of the dynamical system \eqref{eqdiff_pref} can be rewritten as follows:
\[
    \frac{dy_t}{dt} \left[ \frac{1}{y_t}+\frac{1+s}{1-y_t}\right]= \frac{s}{2},
\]
which implies that for any $t\geq0$,
\[\ln(y_t)-(1+s)\ln(1-y_t)=[\ln(b)-(1+s)\ln(1-b)]+\frac{st}{2}.
\]
Therefore for any $t\geq0$,
\[\frac{y_t}{(1-y_t)^{1+s}}=\frac{b}{(1-b)^{1+s}}\exp\left(\frac{st}{2}\right), 
\] which gives the first equation of \eqref{eq-sol-syst-attract}. We now study the two quotients $u/y$ and $v/(1-y)$. From \eqref{eqdiff_pref} we get for all $t\geq 0$,
\begin{equation}\label{u_y_pref}
\begin{aligned} 
\frac{d\left(\frac{u}{y}\right)}{dt}  &=  -\frac{u(1-y)}{y(1+sy)} + \frac{2+s}{4y(1+sy)}(vy + (1-y)u) - \frac{us}{2y}\frac{1-y}{1+sy} \\ & = \frac{(2+s)}{4(1+sy)} (1-y) \frac{v}{1-y} + \frac{u}{y} \biggl[\frac{-(1-y)}{1+sy} + \frac{(2+s)(1-y)}{4(1+sy)} -\frac{s(1-y)}{2(1+sy)}\biggr] \\ & = - \,\frac{(2+s)  (1-y)}{4(1+sy)} \left(\frac{u}{y}- \frac{v}{1-y}\right),
\end{aligned}
\end{equation}
and similarly,
\begin{align} \label{v_1-y_pref}
\frac{d\left(\frac{v}{1-y}\right)}{dt} = \frac{(2+s)  y}{4(1+sy)} \left( \frac{u}{y} - \frac{v}{1-y} \right).
\end{align}
Therefore, denoting for any $t\geq0$, $\Delta_t=\frac{u_t}{y_t}-\frac{v_t}{1-y_t}$ we get for any $t\geq0$,
\begin{align*}
    \frac{d\Delta_t}{dt}
    = - \frac{2+s}{4(1+sy_t)}\Delta_t.
\end{align*}
The first line of Equation \eqref{eqdiff_pref} then gives
$$\frac{1}{\Delta_t}\frac{d\Delta_t}{dt}=-\frac{2+s}{2s}\frac{dy_t}{dt}\left[\frac{1}{y_t}+\frac{1}{1-y_t}\right].$$ Noting that $\Delta_0=1$ and $y_0=b\in(0,1)$, this differential equation can be solved by integrating with respect to $t$, as 
\begin{equation}\label{eq-Delta}
   \Delta_t=\frac{u_t}{y_t}-\frac{v_t}{1-y_t}=\left(\frac{1-y_t}{y_t}\right)^{\frac{2+s}{ 2s}}\left(\frac{b}{1-b}\right)^{\frac{2+s}{2s}}=g(y(t))/g(b)
\end{equation} where $g(y)=\left(\frac{1-y}{y}\right)^{\frac{2+s}{ 2s}}$ which gives the second equation of \eqref{eq-sol-syst-attract}.
Finally, for any $t\geq0$, note that 
\begin{equation}\label{eq-u+v-Delta-y}\frac{d(u_t+v_t)}{dt}=\Delta_t\frac{dy_t}{dt}\end{equation}
Equation \eqref{eq-u+v-Delta-y} is a key relationship between the dynamics of the quantities $y$, $\Delta$ and $u+v$, that is true for all selection models considered so far. It is notably due to the fact that both parents transmit their genome equally to their offspring, independently from their own type and from the type of their offspring.
Using this equation and the time change $t\rightarrow y_t$ gives that \begin{equation}u(t)+v(t)=b+\int_b^{y(t)}\frac{g(w)}{g(b)}dw\end{equation}
which will be true for all selection models considered in this article and ends this proof.
\end{proof}

\subsection{Homogamous choosiness}\label{sec-homo-app}
We now deal with the case where individuals belonging to $\mathcal{Y}_n$ are homogamous with higher probability, whereas individuals from the other type choose their partner uniformly in the population. Explicitly, if the mother belongs to the homogamous type, individuals of that type have a probability $(1+s)$ times higher to be chosen as father than the other. As before, we now give the skeleton of the Markov chain $Y_n$ in this case.

\begin{proposition}\label{prop_Y_Homo}
The stochastic process $(Y_n)_{n\in\mathbb{Z}_+}$ is a Markov chain such that $Y_n\in\{0,...,N\}$ for all $n\in\mathbb{Z_+}$. Moreover,
\begin{itemize}
\item[$(i)$] For any $k\in\{0,...,N\}$ its transition probabilities are given by:
$$\mathbb{P}[Y_{n+1} = k+1 | Y_n = k] =  \dfrac{1}{2}\dfrac{k}{N}\biggl(1-\dfrac{k}{N}\biggr)\biggl[2+\dfrac{k}{N}\biggl(1-\dfrac{k}{N}\biggr)\dfrac{s}{1+s(k/N)}\biggr],$$
\\
     $$\mathbb{P}[Y_{n+1} = k-1 | Y_n = k] = \dfrac{1}{2}\dfrac{k}{N}\biggl(1-\dfrac{k}{N}\biggr)\biggl[2-\biggl(\dfrac{k}{N}\biggr)^2\dfrac{s}{1+s(k/N)}\biggr],$$
    and $\mathbb{P}[Y_{n+1}=k|Y_n =k] = 1 - p_k$, where $p_k = \frac{k}{N}(1-\frac{k}{N})[2 + \frac{s/2}{1+s(k/N)}\frac{k}{N}(1-\frac{2k}{N})]$.
    In particular this Markov chain is absorbed in $0$ and $N$.
\item[$(ii)$] When $N$ goes to infinity, $$\mathbb{P}(T_N<+\infty|Y_0=\lfloor bN\rfloor)\rightarrow 1.$$ 
\end{itemize}
\end{proposition}

\begin{proof} Both points of the proof are slightly more involved than the one in the proof of Proposition \ref{prop_Y_Attract} for the increased attractiveness case. For the first point, from Proposition \ref{prop:dynamics_Z} (or the model description made in Section \ref{sec:model}, $Y_{n+1}=Y_n$ if and only if the replaced individuals $\kappa_n$ is chosen in $I\setminus\mathcal{Y}_n$ and the individual that transmits its type is chosen in $\mathcal{Y}_n$. This gives in this homogamous case that \begin{align*}\mathbb{P}(Y_{n+1}=Y_n+1|Y_n)&=\left(1-\frac{Y_n}{N}\right)\frac{1}{2}\left[\frac{Y_n}{N}+\frac{Y_n}{N}\frac{(1+s)Y_n}{N-Y_n+(1+s)Y_n}+\left(1-\frac{Y_n}{N}\right)\frac{Y_n}{N}\right]\\&=\frac{1}{2}\left(1-\frac{Y_n}{N}\right)\frac{Y_n}{N}\left[2-\frac{Y_n}{N}+\frac{(1+s)Y_n}{N-Y_n+(1+s)Y_n}\right].\end{align*}
The three terms in the right-hand side of the first line of the previous equation respectively correspond to the case where the mother is advantaged and transmits its type (i.e. $\nu_n=\mu_n\in\mathcal{Y}_n$), the case where the father is advantaged and transmits its type while the mother is also advantaged (i.e. $\mu_n,\nu_n=\pi_n\in\mathcal{Y}_n$) and the case where the father is advantaged and transmits its type while the mother is not advantaged.
Similarly, 
\begin{align*}\mathbb{P}(Y_{n+1}=Y_n-1|Y_n)&=\frac{Y_n}{N}\frac{1}{2}\left[\frac{N-Y_n}{N}+\frac{N-Y_n}{N}\frac{N-Y_n}{N}+\frac{Y_n}{N}\frac{N-Y_n}{N-Y_n+(1+s)Y_n}\right]\\&=\frac{1}{2}\left(1-\frac{Y_n}{N}\right)\frac{Y_n}{N}\left[2-\frac{Y_n}{N}+\frac{Y_n}{N-Y_n+(1+s)Y_n}\right].\end{align*}
Therefore
\begin{align*}\mathbb{P}(Y_{n+1}=Y_n+1|Y_{n+1}\neq Y_n,Y_n)&=\frac{\frac{1}{2}\left[2-\frac{Y_n}{N}+\frac{(1+s)Y_n}{N+sY_n}\right]}{2-\frac{Y_n}{N}+\frac{(1+s/2)Y_n}{N+sY_n}}\\&=\frac{1}{2}+\frac{s}{4}\frac{\frac{Y_n}{N}}{2+\frac{5s}{2}\frac{Y_n}{N}-s\left(\frac{Y_n}{N}\right)^2}\\&=\frac{1}{2}+\frac{s}{4}f\left(\frac{Y_n}{N}\right),\end{align*} where $f(y)=\frac{y}{2+\frac{5sy}{2}-sy^2}$ for all $y\in[0,1]$.
For this homogamous case, for any fixed $k\geq 0$, $\mathbb{P}(Y_{n+1}=Y_n+1|Y_{n+1}\neq Y_n,Y_n=k)\rightarrow 1/2$ when $N$ goes to infinity. Therefore the skeleton of the Markov chain $(Y_n)_{n\in\mathbb{Z}_+}$ does not dominate stochastically a biased Markov chain as in the increased attractiveness situation presented and studied in Proposition \ref{prop_Y_Attract}. Nevertheless the function $f$ is decreasing on $[0,1]$ therefore for any $y\in[b/2,1]$, one has $f(y)\geq f(b/2)>0$ and therefore $\mathbb{P}(Y_{n+1}=Y_n+1|Y_{n+1}\neq Y_n,Y_n)\geq \frac{1}{2}+s/4f(b/2)$ as long as $Y_n /N\geq b/2$. Therefore on $[bN/2,N]$, the skeleton of the Markov chain $(Y_n)_{n\in\mathbb{Z}_+}$ dominates stochastically a biased  random walk that jumps from $k$ to $k+1$ with probability $1/2+s/4f(b/2)$ and from $k$ to $k-1$ with probability $1/2-s/4f(b/2)$. Let us denote by $(\underline{Y}_n)_{n\in\mathbb{Z}_+}$ such a random walk and by $T^{\underline{Y}}_y$ its reaching time of $y$ for any $y\in\{0,...,N\}$.
Then \begin{align*}\mathbb{P}(T_0<T_N|Y_0=\lfloor bN\rfloor)&\leq\mathbb{P}(T_{\lfloor bN/2\rfloor}<T_N|Y_0=\lfloor bN\rfloor)\\&\leq\mathbb{P}(T^{\underline{Y}}_{\lfloor bN/2\rfloor}<T^{\underline{Y}}_N|\underline{Y}_0=\lfloor bN\rfloor)\\&\rightarrow 0 \quad\text{when $N$ goes to infinity,} \end{align*}
which ends the proof. The last convergence follows from a classical result on gambler's ruin that can be found in \cite{Feller} for example.
\end{proof}

\begin{proof}[Proof of Theorem \ref{Convergence_homo}] The proof is the same as for Theorem \ref{Convergence_pref}. From Equation \eqref{eq-dynamics} and Proposition \ref{prop_Y_Homo}, one has for all $n\geq0$,
\begin{equation*} 
\begin{aligned} 
\mathbb{E}[Y_{n+1}-Y_n|\mathcal{F}_n^Z]&=\mathbb{P}\left[Y_{n+1}=Y_n+1|\mathcal{F}_n^Z\right]-\mathbb{P}\left[Y_{n+1}=Y_n-1|\mathcal{F}_n^Z\right]\\& = \frac{1}{2}\frac{Y_n}{N}\frac{N-Y_n}{N}\frac{sY_n}{N+sY_n}.\end{aligned}
\end{equation*} Now from Equation \eqref{eq-dynamics} again,
\begin{align*} 
\mathbb{E}[U_{n+1}-U_n|\mathcal{F}_n^Z]&=\mathbb{E}\left[-\sum_{l'\in\mathcal{Y}_0}W_n(\kappa_n,l')\mathbf{1}_{\kappa_n\in\mathcal{Y}_n}+\sum_{l'\in\mathcal{Y}_0}\frac{W_n(\mu_n,l')+W_n(\pi_n,l')}{2}\mathbf{1}_{\nu_n\in\mathcal{Y}_n}\Big|\mathcal{F}_n^Z\right]\\& = -\frac{Y_n}{N}\frac{U_n}{Y_n} +\mathbb{E}\left[\mathbf{1}_{\nu_n=\mu_n}\mathbf{1}_{\mu_n\in\mathcal{Y}_n}\sum_{l'\in\mathcal{Y}_0}\frac{W_n(\mu_n,l')+W_n(\pi_n,l')}{2}\right.\\&\phantom{-\frac{Y_n}{N}\frac{U_n}{Y_n} +\mathbb{E}}\quad\left.+\mathbf{1}_{\nu_n=\pi_n}\mathbf{1}_{\pi_n\in\mathcal{Y}_n}\sum_{l'\in\mathcal{Y}_0}\frac{W_n(\mu_n,l')+W_n(\pi_n,l')}{2}\right]\\&=-\frac{U_n}{N}+\frac{1}{4}\frac{Y_n}{N}\left[\frac{U_n}{Y_n}+\frac{(1+s)Y_n}{(1+s)Y_n+N-Y_n}\frac{U_n}{Y_n}\right.\\&\phantom{-\frac{U_n}{N}+\frac{1}{4}\frac{Y_n}{N}[]}\left.+\frac{N-Y_n}{(1+s)Y_n+N-Y_n}\frac{V_n}{N-Y_n}\right]\\&\phantom{=-\frac{U_n}{N}}+\frac{1}{4}\frac{Y_n}{N}\left[\frac{(1+s)Y_n}{(1+s)Y_n+N-Y_n}\frac{2U_n}{N}+\frac{N-Y_n}{N}\left(\frac{V_n}{N-Y_n}+\frac{U_n}{Y_n}\right)\right]\\&=-\frac{U_n}{2N}+\frac{3(1+s)Y_nU_n}{4N(N+sY_n)}+\frac{Y_nV_n}{4N(N+sY_n)}\\&+\frac{Y_nV_n}{4N(N+sY_n)}+\frac{Y_nV_n}{4N^2}-\frac{Y_nU_n}{4N^2}\\&=\frac{-U_n(N-Y_n)}{N+sY_n}+\frac{(2N+sY_n)(Y_nV_n+(N-Y_n)U_n)}{4(N+sY_n)}.
\end{align*}
Similarly,
\begin{align*}
\mathbb{E}[V_{n+1}-V_n|\mathcal{F}^Z_n]&=\mathbb{E}\left[-\sum_{l'\in\mathcal{Y}_0}W_n(\kappa_n,l')\mathbf{1}_{\kappa_n\notin\mathcal{Y}_n}+\sum_{l'\in\mathcal{Y}_0}\frac{W_n(\mu_n,l')+W_n(\pi_n,l')}{2}\mathbf{1}_{\nu_n\notin\mathcal{Y}_n}\Big|\mathcal{F}_n^Z\right]\\& = -\frac{N-Y_n}{N}\frac{V_n}{N-Y_n} +\mathbb{E}\left[\mathbf{1}_{\nu_n=\mu_n}\mathbf{1}_{\mu_n\notin\mathcal{Y}_n}\sum_{l'\in\mathcal{Y}_0}\frac{W_n(\mu_n,l')+W_n(\pi_n,l')}{2}\right.\\&\phantom{-\frac{N-Y_n}{N}\frac{V_n}{N-Y_n} +\mathbb{E}}\quad\left.+\mathbf{1}_{\nu_n=\pi_n}\mathbf{1}_{\pi_n\notin\mathcal{Y}_n}\sum_{l'\in\mathcal{Y}_0}\frac{W_n(\mu_n,l')+W_n(\pi_n,l')}{2}\right]\\&=-\frac{V_n}{N}+\frac{1}{4}\frac{N-Y_n}{N}\left[\frac{V_n}{N-Y_n}+\frac{Y_n}{N}\frac{U_n}{Y_n}+\frac{N-Y_n}{N}\frac{V_n}{N-Y_n}\right]\\&\phantom{=-\frac{V_n}{N}}+\frac{1}{4}\frac{N-Y_n}{N}\left[\frac{N-Y_n}{N}\frac{2V_n}{N-Y_n}\right.\\&\phantom{=-\frac{V_n}{N}+\frac{1}{4}\frac{N-Y_n}{N}[\quad}\left.+\frac{Y_n}{(1+s)Y_n+N-Y_n}\left(\frac{U_n}{Y_n}+\frac{V_n}{N-Y_n}\right)\right]\\&=-V_nY_n+\frac{(2N+sY_n)[(N-Y_n)U_n+Y_nV_n]}{4N^2(N+sY_n)}.
\end{align*}
Therefore, recalling that $Z_n=\frac{1}{N}(Y_n,U_n,V_n)$, we get:
$$\mathbb{E}(Z_{n+1}-Z_n|\mathcal{F}_n^Z)=\frac{1}{N}g(Z_n),$$
where \begin{align*}g(y,u,v)&=\Big(\frac{1}{2}\frac{sy^2(1-y)}{1+sy},\\&\quad\quad-\frac{u(1-y)}{1+sy}+ \frac{1}{4}\frac{2+sy}{1+sy}(vy + (1-y)u),\\&\quad\quad-vy+ \frac{1}{4}\frac{2+sy}{1+sy}(vy+ (1-y)u)\Big).\end{align*}
This expression for $\mathbb{E}(Z_{n+1}-Z_n|\mathcal{F}_n^Z)$ allows us to get an analogous decomposition to the one in Theorem \ref{Convergence_pref}. As the Jacobian of $g$ is bounded here too, we can conclude following the same steps.
\end{proof}

\begin{proof}[Proof of Proposition \ref{Solution_eqdiff_homo}] The existence and uniqueness of a solution of the differential system \eqref{eqdiff_homo} follows from Cauchy-Lipschitz theorem. This theorem also gives that any solution starting from $y_t=b\in(0,1)$ is such that $y_t\in(0,1)$ for all $t\geq 0$. Now as in the proof of Proposition \ref{Solution_eqdiff_pref}, the dynamics of the number of advantaged individuals given in the first equation of System \eqref{eqdiff_homo} is independent from the two other equations, and can be rewritten as
    \begin{equation*}
        \frac{dy_t}{dt} \left[ \frac{1+s}{y_t}+ \frac{1}{y_t^2} +\frac{1+s}{1-y_t}\right]= \frac{dy_t}{dt}\frac{d \,\log (\rho (y_t))}{dy_t} = \frac{s}{2},
    \end{equation*}
    where $\rho(y)=\exp(-1/y)(\frac{y}{1-y})^{(1+s)}>0$ for all $y\in(0,1)$. Therefore
    $$y_t=\rho^{-1}\left(\rho(b)\exp\left(\frac{s}{2}t\right)\right), \; \; t \geq 0.$$
    
Let us now compute, as in the proof of Proposition \ref{Solution_eqdiff_pref}, the derivative with time of the averaged weights $u_t/y_t$ and $v_t/(1-y_t)$. Firstly,
\begin{equation*}
\begin{aligned} 
\biggl(\frac{u}{y}\biggr)^{'} &= \frac{u'}{y}-u\frac{y'}{y^2} \\[5pt]
&= -\frac{u}{y}\frac{1-y}{1+sy}+\frac{1}{4y}\frac{2+sy}{1+sy}(u(1-y)+vy) - u\frac{(s/2)(1-y)}{1+sy} \\[5pt]
&= \frac{1}{1+sy}\Big[\;\frac{u}{y}\Big(-(1-y)+\frac{2+sy}{4}(1-y)-\frac{s}{2}y(1-y)\Big) +\frac{v}{1-y}\Big(\frac{2+sy}{4}(1-y)\Big)\Big] \\[5pt]
&= -\frac{1}{4}\biggl(\frac{2+sy}{1+sy}\biggl)(1-y)\Big(\frac{u}{y}-\frac{v}{1-y}\Big).
\end{aligned}
\end{equation*}
Similarly, 
\begin{equation}\label{v_1-y_homo}
\begin{aligned}
\biggl(\frac{v}{1-y}\biggr)^{'} &= \frac{v'}{1-y}+v\frac{y'}{(1-y)^2} \\[5pt]
&= -\frac{vy}{1-y}+\frac{1}{4(1-y)}\biggl(\frac{2+sy}{1+sy}\biggl)(u(1-y)+vy) - v\frac{(s/2)y^2}{(1-y)(1+sy)} \\[5pt]
&=  \frac{1}{1+sy}\Big[\;\frac{u}{y}\frac{y(2+sy)}{4y} +\frac{v}{1-y}\Big(-y(1+sy) + \frac{2+sy}{4}y + y^2(s/2) \Big)\Big]\\[5pt]
&= \frac{1}{4} \biggl(\frac{2+sy}{1+sy}\biggl)\,y\,\Big(\frac{u}{y}-\frac{v}{1-y}\Big).
\end{aligned}
\end{equation}
Subtracting these two expressions gives
\begin{equation*}
    \Delta_t' := \left(\frac{u}{y}-\frac{v}{1-y}\right)' =  -\frac{1}{4}\Big(\frac{2+sy}{1+sy}\Big)\Big(\frac{u}{y}-\frac{v}{1-y}\Big)= -\frac{1}{4}\Big(\frac{2+sy}{1+sy}\Big)\Delta_t.
\end{equation*}
Using the change of variables $y(x)=w$ and the fact that $\Delta_0 = 1$, we can solve the previous ordinary differential equation:
\begin{equation}\label{eqDeltah}
\begin{aligned}
    \frac{u_t}{y_t}-\frac{v_t}{1-y_t}=\Delta_t &= \exp\Big( -\frac{1}{4}\int_0^t\frac{2+sy_x}{1+sy_x}dx\Big) \\
    &= \exp\Big( -\frac{1}{2s}\int_0^t\frac{y_x'(2+sy_x)}{y_x^2 (1-y_x)}dx \Big) \\
    &= \exp\Big(- \frac{1}{2s}\int_0^t\Big[\frac{(2+s)y_x'}{y_x}+\frac{(2+s)y_x'}{1-y_x}+\frac{2y_x'}{y_x^2}\Big]dx \\
    &= \exp\Big(-\frac{2+s}{2s}\log\frac{(1-b)y_t}{(1-y_t)b}\Big)\exp\Big(\frac{1}{sy_t}-\frac{1}{sb}\Big)\\&=\left(\frac{b}{1-b}\right)^{\frac{2+s}{2s}}\left(\frac{1-y_t}{y_t}\right)^{\frac{2+s}{2s}}\exp\Big(\frac{1}{sy_t}-\frac{1}{sb}\Big)=g(y_t)/g(b)
\end{aligned}
\end{equation}
where $g(y)=\left(\frac{1-y_t}{y_t}\right)^{\frac{2+s}{2s}}\exp\Big(\frac{1}{sy_t}\Big)$.
Finally as in the proof of Proposition \ref{Solution_eqdiff_pref}, Equation \eqref{eq-u+v-Delta-y} is satisfied which ends the proof similarly.
\end{proof}

\bibliographystyle{elsarticle-harv}
\bibliography{biblio}

\end{document}